\newif\ifRelease
\Releasetrue
\ifRelease
 \newcommand{\layout}{twocolumn}
\else
 \newcommand{\layout}{onecolumn}
\fi
\documentclass[journal=mamobx,manuscript=letter,layout=\layout]{achemso}
\SectionNumbersOn

\usepackage[english]{babel}
\usepackage[T1]{fontenc}
\usepackage{ae}
\usepackage{aecompl}

\usepackage{iftex}
\ifLuaTeX
 \usepackage[utf8]{luainputenc}
\else
 \usepackage[utf8]{inputenc}
\fi

\DeclareUnicodeCharacter{22C5}{$\cdot$}
\DeclareUnicodeCharacter{1D712}{$\chi$}
\DeclareUnicodeCharacter{226}{\AA}

\usepackage{color}

\usepackage{graphicx}
\usepackage{float}

\usepackage{tabularx}
\usepackage{booktabs}

\usepackage{siunitx}
\usepackage[version=4]{mhchem}

\usepackage{xr}
\usepackage{hyperref}
\usepackage[capitalise,noabbrev]{cleveref}

\newcommand*{\citen}[1]{%
  \begingroup
    \romannumeral-`\x %
    \setcitestyle{numbers}%
    \cite{#1}%
  \endgroup
}

\newcommand{\sysTwenty}{20:1:0}
\newcommand{\sysTen}{20:2:1}
\newcommand{\sysFive}{10:2:1}
\newcommand{\sysTwo}{4:2:1}
\newcommand{\sysOne}{2:2:1}
\newcommand{\sysOneThfA}{5:5:3}
\newcommand{\sysOneThfB}{4:4:3}
\newcommand{\sysOneGroup}{$x$:$x$:$y$}
\newcommand{\sysEqui}{3:2:1}

\newcommand{\paperTitle}{Understanding the Lithium Ion Transport in Concentrated Block--Copolymer Electrolytes on a Microscopic Level}

\author{Len Kimms}
\email{len.kimms@uni-muenster.de}
\author{Diddo Diddens}
\email{d.diddens@fz-juelich.de}
\altaffiliation{Current address: Helmholtz Institute M{\"u}nster (IMD-4), Ionics in Energy Storage, Forschungszentrum J{\"u}lich GmbH, Corrensstra{\ss}e 46, 48149 M{\"u}nster, Germany}
\author{Andreas Heuer}
\email{andheuer@uni-muenster.de}
\affiliation{Institut für physikalische Chemie, Universität Münster, Corrensstraße 28/30, 48149 Münster, Germany}

\title{\paperTitle}

\begin{document}

\maketitle

\begin{abstract}
Block--copolymer electrolytes with lamellar microstructure show promising results regarding the ion transport in experiments.
Motivated by these observations we study block--copolymers consisting of a polystyrene (PS) block and a poly(ethylene oxide) (PEO) block which were assembled in a lamellar structure.
The lamella was doped with various amounts of lithium-bis(trifluoromethane)sulfonimide (LiTFSI) until very high loadings with ratios of EO monomers to cations up to 1:1 were reached.
We present insights into the structure and ion transport from extensive Molecular Dynamics simulations.
For high salt concentrations most cations are not coordinated by PEO but rather by TFSI and THF.
More specifically, LiTFSI partially separates from the PEO domain and forms a network--like structure in the middle of the lamella.
This central salt--rich layer plays a decisive role to enable remarkably good cationic mobilities as well as high transport numbers in agreement with the experimental results.
\end{abstract}

\section{Introduction}
Solid polymer electrolytes (SPEs) are extensively investigated because of their potential to unlock novel battery chemistry.
\cite{diasTrendsPolymerElectrolytes2000, fergusCeramicPolymericSolid2010, ngaiReviewPolymerElectrolytes2016, plackeLithiumIonLithium2017}
Replacing the commonly used graphite anode material with lithium metal gives rise to batteries with higher capacities and higher specific energies.
\cite{xuLithiumMetalAnodes2014, plackeLithiumIonLithium2017, linRevivingLithiumMetal2017}
Such secondary lithium metal batteries suffer from electrolyte depletion due to parasitic side reactions and possibly fatal thermal runaway due to dendrite growth.
\cite{rossoOnsetDendriticGrowth2001, aurbachShortReviewFailure2002, biekerElectrochemicalSituInvestigations2015, chengReviewSolidElectrolyte2016}
Electrolytes consisting of flammable organic liquids do not prevent parasitic side reactions and dendrite formation but offer good ionic conductivities.
\cite{xuNonaqueousLiquidElectrolytes2004, sasakiOrganicElectrolytesSecondary2008, dengLiionBatteriesBasics2015}
SPEs are less volatile and therefore increase cell safety.

Poly(ethylene oxide) (PEO) has been identified as a suitable candidate for SPEs.
\cite{wrightElectricalConductivityIonic1975, berthierMicroscopicInvestigationIonic1983, armandPolymerElectrolytes1986, armandPolymersIonicConductivity1990}
Unfortunately PEO exhibits poor ionic conductivity at ambient temperature caused by crystallization.
The conductivity of PEO has a low cationic contribution because the polymer chains strongly coordinate the cations.
\cite{berthierMicroscopicInvestigationIonic1983, armandPolymerElectrolytes1986, donosoNuclearMagneticRelaxation1993}
It has been shown that the conductivity can be improved by increasing the segmental motion of the polymer chains.
\cite{watanabeIonicConductivityMobility1985, donosoNuclearMagneticRelaxation1993, shiEffectMolecularWeight1993}
Therefore, shorter PEO chains are favored because they have a higher mobility and tend to crystallize less.
On the other hand Monroe and Newman \cite{monroeDendriteGrowthLithium2003, monroeEffectInterfacialDeformation2004, monroeImpactElasticDeformation2005}
have shown that a high shear modulus is essential in order to prevent dendrite formation.
The necessary rigidity is hardly reachable even with long PEO chains and hinders the ion transport.\cite{youngMixedSaltEffectsIonic2011, youngBlockCopolymerElectrolytes2014, pandayEffectMolecularWeight2009}
A proposed solution to the conflicting mechanical and electrical properties are block copolymers (BCPs) which are able to decouple both properties as follows:
BCPs consist of chemically dissimilar polymer blocks that are covalently bonded end--to--end so that each block can be optimized for a specific property.
Typically, nonconductive blocks are used to tune mechanical properties and PEO is used to form conducting pathways.
\cite{sooRubberyBlockCopolymer1999, niitaniSynthesisLiIon2005, singhEffectMolecularWeight2007, hallinanLithiumMetalStability2013, youngBlockCopolymerElectrolytes2014}
Depending on the properties of the individual blocks BCPs give rise to microphase separation and the conductivity consequently depends on the formed microstructure.
\cite{pandayEffectMolecularWeight2009, matsenBlockCopolymerMicrostructures1997, youngBlockCopolymerElectrolytes2014}
Contrary to PEO homopolymers the conductivity of such BCPs increases with longer PEO chains for intermediate chain lengths.\cite{singhEffectMolecularWeight2007, pandayEffectMolecularWeight2009, ganesanMechanismsUnderlyingIon2012}

Dörr et al.~\cite{dorrAmbientTemperatureElectrolyte2018} and Pelz et al.~\cite{pelzSelfAssembledBlockCopolymer2019}
have studied block copolymers with unusually short PEO chains ($M_W = \SI{2000}{\gram\per\mol}$) that allow high lithium salt loadings of \ce{LiTFSI}.
In the experiments a ratio of 4.3 \ce{Li+} per ethylene oxide monomer (EO) has been identified as optimum in the conductivity.
Those BCPs self--assemble into lamellar microstructures that exhibit a remarkably good total conductivity of \SI{2.2}{\milli\siemens\per\centi\meter} at \SI{20}{\celsius} with a weak temperature dependence
and a high lithium ion transference number of 0.7.\cite{pelzSelfAssembledBlockCopolymer2019}
The preparation was done by evaporation--induced self--assembly using tetrahydrofuran (\ce{THF}) as solvent.
\ce{THF} could not be evaporated completely from these samples
and subsequent experimental investigations indicate that \ce{THF} takes an important role in the coordination of \ce{Li+}.\cite{dorrAmbientTemperatureElectrolyte2018,zhou_deciphering_2021,krause_superionic_2024}
It was found that thermal annealing increases the conductivity presumably by a long--range ordering of the lamellar structures.\cite{dorrAmbientTemperatureElectrolyte2018}
Utilizing small--angle X--ray scattering (SAXS) it has been observed that the addition of salt selectively swells the PEO domain
which results in an increase of the overall bilayer thickness from \SI{28}{\nano\meter} in neat polymer to \SI{52}{\nano\meter} in the best conducting sample.\cite{dorrAmbientTemperatureElectrolyte2018}
This pronounced swelling can be an indication of the formation of a salt--rich phase in the PEO domain.
Remarkably wide--angle X--ray scattering (WAXS) reveals a strongly crystalline phase that did not vanish even at a high temperature of \SI{120}{\celsius}.\cite{pelzSelfAssembledBlockCopolymer2019}
Recent investigations have shown high ionic conductivities in other microstructures as well and postulated the presence of a salt--rich phase with additional \ce{THF} as solvent.\cite{krause_superionic_2024}

In this contribution we focus on the \ce{Li+} coordination in the high performance electrolyte with a lamellar microstructure
and provide insights into the structure and dynamics of the reported lamellae by means of molecular dynamics (MD) simulations.
Specifically, we show that a salt--rich central layer forms in systems with high salt loadings
and that this layer is the origin of the desirable electrochemical properties, namely enhanced cation transport and a high cation transference number.
We can consequently rationalize the experimentally observed salt--rich phase, the high cationic transference number, and the swelling of the PEO domain.
These findings can also be applied to other microstructures with high salt loadings, as the ion transport is decoupled from the polymer host.
The next section presents details of the simulation method, followed by a discussion of our results.
In the end we give a brief summary of our findings.

\section{\label{sec:sim_details}Simulation Details}
\begin{figure*}
 \centering
 \includegraphics[width=0.8\textwidth]{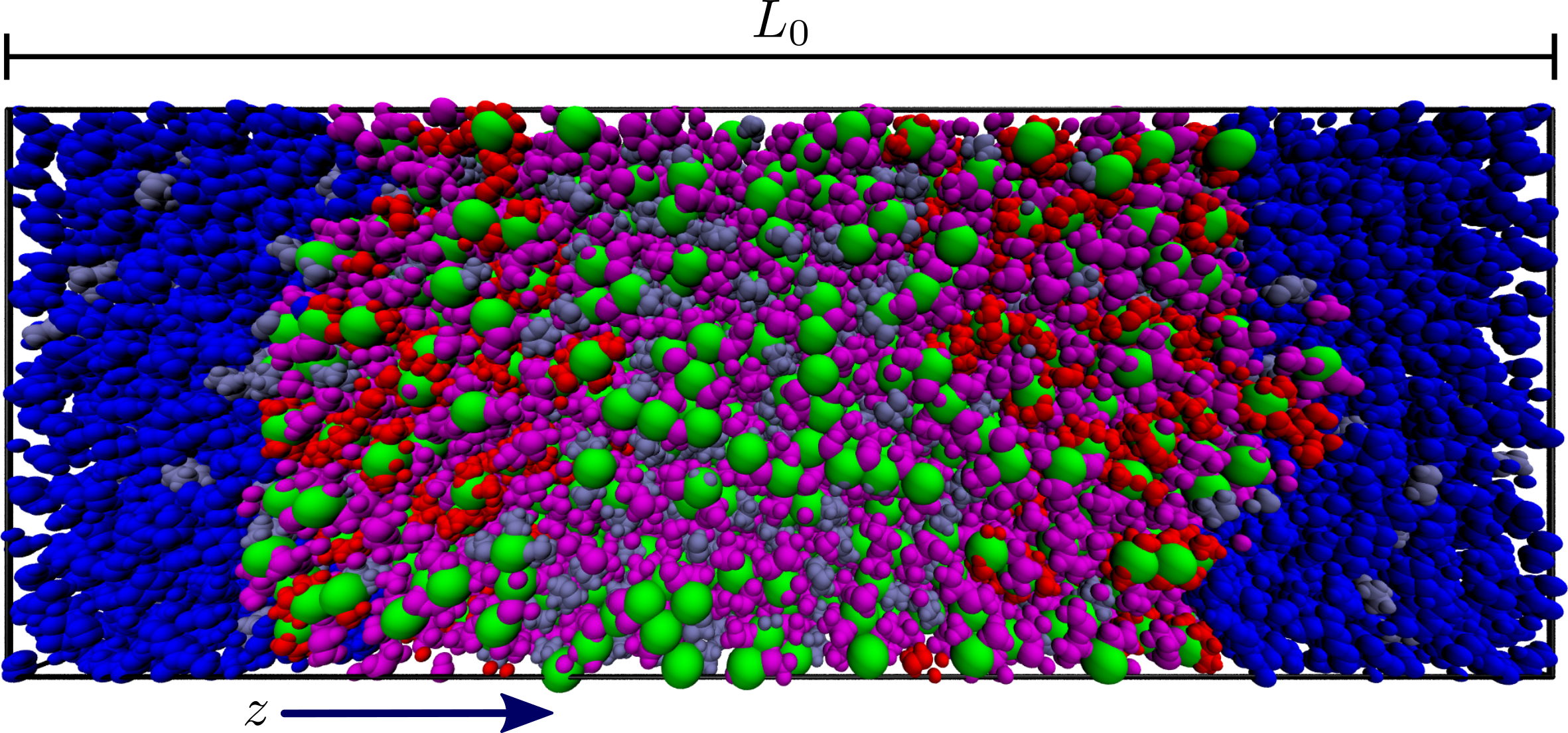}
 \caption{\label{fig:snapshot}Snapshot of the \sysOne{} system with bilayer height $L_0$: PS blue, PEO red, \ce{Li+} green, \ce{TFSI} magenta, \ce{THF} iceblue.}
\end{figure*}

Version 2019 of the molecular dynamics simulation package GROMACS \cite{abrahamGROMACSHighPerformance2015} was used to perform all simulations within a three--dimensional periodic cubic simulation box.
Each simulation box contains a certain number of polymer chains, \ce{LiTFSI} ion pairs and \ce{THF}.
An overview of the composition of the simulated systems is given in \cref{tab:systems}.
\begin{table*}
 \caption{\label{tab:systems}Composition of the simulated systems.}
 \begin{tabularx}{\textwidth}{@{}>{\centering\arraybackslash}X>{\centering\arraybackslash}X>{\centering\arraybackslash}X>{\centering\arraybackslash}X@{}}
  \toprule
   $\text{EO} : \ce{Li+} : \text{THF}$ & $N_{\text{Polymer}}$ & $N_{\text{LiTFSI}}$ & $N_{\text{THF}}$ \\ \hline
   \sysTwenty & 50 & 120 & --- \\
   \sysTen & 50 & 240 & 120 \\
   \sysFive & 50 & 480 & 240 \\
   \sysTwo & 50 & 1200 & 600 \\ \hline
   \sysEqui & 50 & 1600 & 800 \\ \hline
   \sysOne & 50 & 2400 & 1200 \\
   \sysOneThfA & 50 & 2400 & 1440 \\
   \sysOneThfB & 50 & 2400 & 1800 \\
  \bottomrule
 \end{tabularx}
\end{table*}
The systems are named from here on according to their ratio of EO to cations to \ce{THF}.
Although experimentally systems with a ratio of EO to \ce{Li+} of up to 1:5 have been investigated in references \citen{dorrAmbientTemperatureElectrolyte2018, pelzSelfAssembledBlockCopolymer2019}
the following simulations only consider ratios of up to 1:1.
This was done to decrease the computational cost and because higher salt concentrations would only provide limited additional insights into the electrolyte structure.
As discussed later, it can be expected that additional salt only thickens the salt layer in the middle rather than interacting with the block copolymer.
Generally the number of \ce{THF} molecules is chosen to be half the number of ion pairs (see discussion in SI).
It should be noted that the \sysTwenty{} system does not contain any \ce{THF} and serves as a reference system with a typical lithium concentration.
Other exceptions are the \sysOneThfA{} and \sysOneThfB{} systems that contain \SI{20}{\percent} and \SI{50}{\percent} more \ce{THF} as compared to the \sysOne{} system.
This is due to the uncertainty in the experimental measurements of the \ce{THF} content and to test the influence of \ce{THF} in those salt--rich systems.
The \sysOne{}, \sysOneThfA{}, and \sysOneThfB{} systems have a ratio of EO to cations of 1:1 and are therefore collectively referred to as \sysOneGroup{} systems.
All but the \sysOneGroup{} systems have been prepared the following way (cf.~\cref{fig:snapshot} and \cref{fig:si_snapshot} in SI):
The polymers are copolymers consisting of a PS block with 48 monomers and a PEO block with likewise 48 monomers.
Those polymers have the same number of EO monomers as in the experiments done by Dörr et al.~\cite{dorrAmbientTemperatureElectrolyte2018} and Pelz et al.\cite{pelzSelfAssembledBlockCopolymer2019}
In order to reduce the computational cost the length of the PS block is shorter than in these experiments.
Furthermore, there is no third polyisoprene block like in the experimental samples.
However, these changes are inconsequential for the structure of the ion conductive domain, as these blocks are apolar and hence are not expected to interact with the salt (see also below).
PACKMOL \cite{martinezPACKMOLPackageBuilding2009} was used to generate the starting configurations.
The block copolymers are initially arranged in a bilayer so that the same blocks of the bottom and top layer point towards each other and form a PEO domain and a PS domain.
There is one bilayer within the simulation box and \ce{THF} is randomly placed across the whole box.
The ion pairs are randomly positioned within the PEO domain because it has been shown that ions preferentially reside in this polar oxygen--rich domain.
\cite{chuIonDistributionMicrophaseSeparated2018, gilbertDeterminationLithiumIonDistributions2015}
The \sysOneGroup{} systems were prepared based on the structure of the \sysTwo{} system after equilibration.
Additional \ce{LiTFSI} and \ce{THF} was inserted in the middle of the PEO domain and the equilibration routine was repeated.
This preparation procedure is justified by computational and experimental arguments in \cref{sec:comp_layer}.
Simulations were performed using a \textit{NpT} ensemble controlled
by a Nosé--Hoover thermostat \cite{noseUnifiedFormulationConstant1984, hooverCanonicalDynamicsEquilibrium1985} and Parrinello--Rahman barostat.\cite{parrinelloPolymorphicTransitionsSingle1981}
The barostat acts semiisotropically with a coupling pressure of \SI{1}{\bar}.
A \SI{50}{\nano\second} equilibration run was performed with $T = \SI{450}{\kelvin}$ after the initial simulation box was shrunken to a reasonable size.
Subsequently, the temperature was reduced to $T = \SI{400}{\kelvin}$ and the systems were equilibrated for at least \SI{500}{\nano\second}.
Afterwards a production run of \SI{500}{\nano\second} was performed for all systems but the \sysEqui{} system,
which was only used to study the partial separation of salt from the polymer domain (see below).
Simulations were propagated with a time step of \SI{2}{\femto\second} and a cutoff of \SI{1.4}{\nano\meter} for Van der Waals interactions.
The Coulomb interactions were treated with the particle mesh Ewald (PME) method.\cite{dardenParticleMeshEwald1993, essmannSmoothParticleMesh1995}
A cutoff distance of \SI{1.4}{\nano\meter}, a grid spacing of \SI{0.1}{\nano\meter}, and an interpolation order of 6 are used as PME settings.

To describe the particle interactions the OPLS \cite{jorgensenDevelopmentTestingOPLS1996} force field was used.
The \ce{Li+} particles are described by parameters from reference \citen{aqvistIonwaterInteractionPotentials1990}
and the parameters of PEO and \ce{THF} are chosen according to reference \citen{mongelliCriticalPolyalkanePolyether2016}.
The force field parameters of PS are based on references \citen{lemkulPracticalConsiderationsBuilding2010, torres-knoopEntropicSeparationStyrene2015}.
TFSI anions are parameterized as described in reference \citen{gouveiaIonicLiquidsAnions2017}.
The OPLS force field does not account for polarization effects.
To approximate polarization a factor of 0.8 is used to scale partial charges of both ion species.
Several other studies have shown that this leads to better agreement with experiments.\cite{leontyevElectronicContinuumModel2010, shimizuStructuralAggregateAnalyses2015, thumSolvateIonicLiquids2020}

\section{Results and Discussion}

\subsection{\label{sec:comp_layer}Composition Across the Bilayer}
\begin{figure*}
 \centering
 \includegraphics{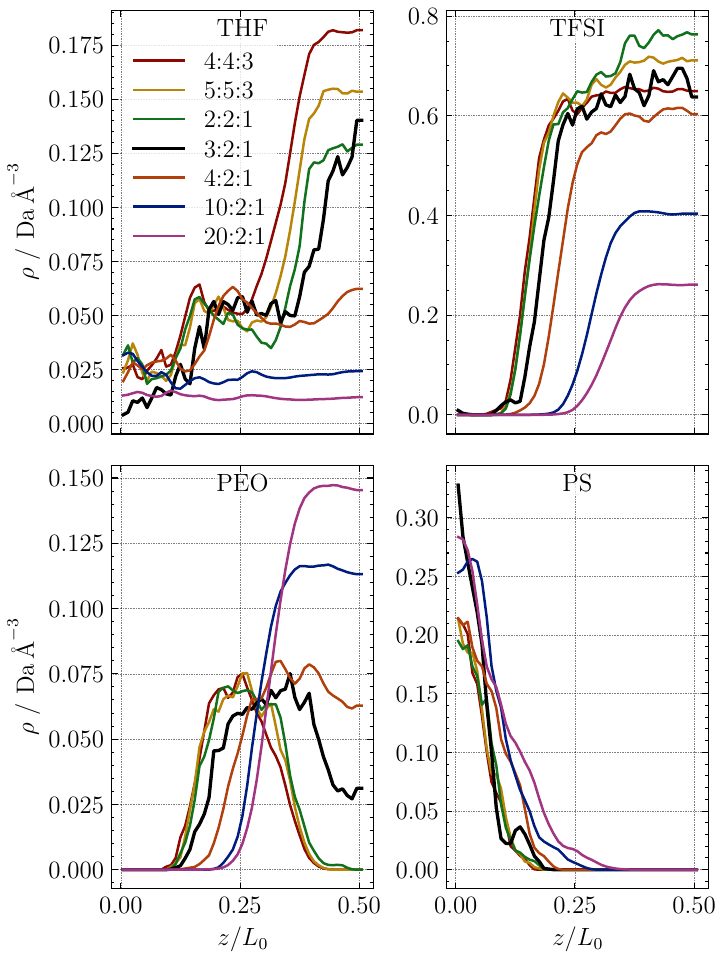}
 \caption{\label{fig:res_profile}Distribution of the mass density $\rho$ of the different residues along the normalized bilayer height $z/L_0$.
 \ce{Li+} correlates with \ce{TFSI} and is not shown. Due to the symmetry of the bilayer only one half of the density profile is shown.
 The densities are averaged over the whole production run except the \sysEqui{} system which is an average of the last \SI{50}{\nano\second} of the equilibration run (see \cref{sec:sim_details}).}
\end{figure*}

First the distribution of the different components is analyzed with respect to the bilayer normal $z$ (see \cref{fig:snapshot}).
In the following the distribution of the mass density $\rho$ of a certain component is calculated along the normalized bilayer height $z/L_0$.
Absolute values for the bilayer height $L_0$ are tabulated in SI \cref{tab:box_dims}.
Because of the symmetry of the bilayer, the density in the interval $z/L_0 \in [0, 0.5)$ is mirrored in the interval $z/L_0 \in [0.5, 1)$.
For this reason the mass density $\rho$ is averaged over both intervals and subsequent depictions show solely the interval $z/L_0 \in [0, 0.5)$.
An overview over the densities of all residues is given in \cref{fig:res_profile}.
During the extensive equilibration and production run the \ce{LiTFSI} salt exclusively stays within the PEO domain.
This matches the assumed behavior during construction of the initial configuration in \cref{sec:sim_details}.
Other simulation studies have observed a small population of ions which reside within the nonconductive PS phase, which exhibit stronger anion--cation coordination and lower mobilities,
when using \ce{LiPF6} as salt and a different force field.\cite{sethuramanMultiscaleSimulationsLamellar2017, sethuramanIonTransportMechanisms2017, zhangMechanismsIonTransport2019}
Apparently, \ce{LiTFSI} has a lower tendency to form small neutral clusters that can migrate into the apolar PS domain because this salt dissociates more easily.

\begin{figure}
 \centering
 \includegraphics[width=0.4\textwidth]{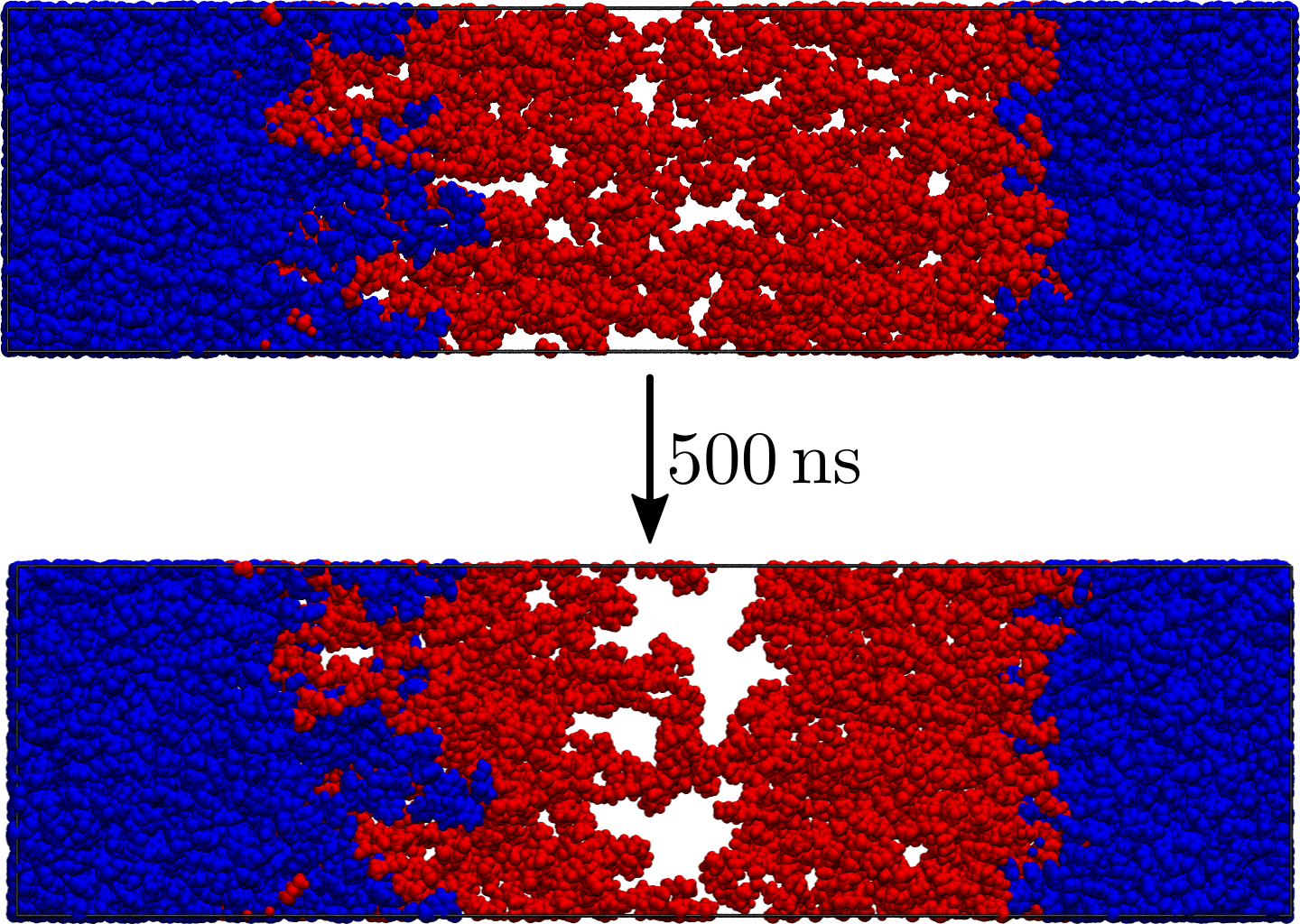}
 \caption{\label{fig:snapshot_3to2to1}Snapshots of the initial polymer structure and the polymer localization after \SI{500}{\nano\second} of equilibration in the \sysEqui{} system.
 \ce{LiTFSI} and \ce{THF} are not shown to highlight the formation of the salt layer in the middle.
 The accompanying volume change during the simulation is too small to be clearly visible in the snapshots.}
\end{figure}

The distribution of the mass density $\rho$ of PEO in the \sysTwo{} system decreases towards the middle of the bilayer at $z/L_0 = 0.5$
(note that this also allowed for the insertion of additional salt to create the \sysOneGroup{} systems, see \cref{sec:sim_details}).
The decrease of PEO towards the middle of the bilayer is even more pronounced in the \sysEqui{} system, which was deliberately set up to study the formation of the salt--rich layer.
\Cref{fig:snapshot_3to2to1} allows for a visual inspection of the polymer chains in this system which clearly shows a separation of the bilayer in the middle during equilibration.
This emerging gap is occupied by a mixture of \ce{THF} and \ce{LiTFSI}.
As indicated by this observation, it is reasonable to assume that a salt layer is formed in highly concentrated systems.
The separation of the bilayer is traceable by observing the distance between the polymer ends as shown in \cref{fig:si_nonequi_end_dist_middle_ts}.
However, the equilibration run of the \sysEqui{} system shows that the formation of the salt layer is barely completed even after \SI{1}{\micro\second} of simulation.
In this system the salt gradually leaves the PEO domain which causes an increase of PEO in that region (see \cref{fig:si_peo_profile_ts,fig:si_peo_p_ts}).
This suggests a rather slow assembly of the salt layer and justifies our strategy to insert this layer directly in the \sysOneGroup{} systems which yields equilibrated structures at a significantly reduced computational cost,
as in the \sysOneGroup{} systems the salt layer remains present over the whole simulation time.
Analysis of the diffusion coefficients within this salt layer in \cref{sec:cond} shows that all electrolyte components exhibit displacements during the production run
that are sufficiently large to allow exchange between the salt layer and the PEO domain.
As discussed in \cref{sec:sim_details}, this salt layer has also been observed experimentally, which also motivated our study.\cite{dorrAmbientTemperatureElectrolyte2018, pelzSelfAssembledBlockCopolymer2019}

Of particular interest is the distribution of \ce{THF} because of its postulated important role in the coordination of \ce{Li+} in reference \citen{dorrAmbientTemperatureElectrolyte2018}.
The mass density $\rho$ of \ce{THF} along the normalized bilayer height $z/L_0$ is shown in \cref{fig:res_profile} (upper left).
The \sysTen{} system exhibits an approximately homogeneous distribution of \ce{THF} across the PS and PEO block.
\ce{THF} acts in this case as a non--selective solvent that has no preference for one block over the other.
In the \sysFive{} system \ce{THF} is most concentrated between the adjacent PS domains of the top and bottom layer around $z/L_0 = 0$.
For higher salt concentrations however \ce{THF} prefers residence in the polar salt--rich region due to favorable interactions with \ce{LiTFSI}.
A large proportion of \ce{THF} is found in the PEO--depleted middle layer where the stoichiometric ratio of [\ce{Li+}:\ce{THF}] can be up to $\approx 1$ (cf.~\cref{tab:si_systems_salt_only} in SI).
Similar ratios have also been found experimentally.\cite{zhou_deciphering_2021,krause_superionic_2024}
This relative amount of \ce{THF} is largest for a very small total amount of \ce{THF} (see \cref{fig:si_conc_profile_li_thf} in SI),
which indicates that \ce{THF} preferentially compensates the missing PEO coordination of \ce{Li+}.
Interestingly there is a locally higher density of \ce{THF} at the interface between PS and PEO blocks of the block copolymers.
This effect has been attributed to a decrease in contacts between the immiscible PS and PEO blocks caused by the solvent.\cite{lifschitzInterfacialBehaviorCompressible1993, huangAdsorptionMinorityComponent1996}
An orientation in which the \ce{THF} oxygens point towards the PEO domain is favored at this interface (not shown).
The shift of this locally higher density of \ce{THF} between the \sysTwo{} and \sysOne{} systems hints at a swelling of the PEO domain due to the additional salt.
Motivated by the observation that \ce{THF} preferentially resides in the salt rich domain the coordination environment of \ce{Li+} is further characterized in the following sections.

\subsection{Coordination Numbers}
Radial distribution functions (RDFs) are used to identify atoms that coordinate cations.
The examination of the RDFs of different oxygen types, belonging to anions, PEO, and THF
shows a distinct inner peak for all three residues that is caused by the first coordination shell (see SI \cref{fig:si_rdf_tfsi,fig:si_rdf_ether}).
The shape of the RDFs has been extensively discussed in the past.\cite{borodinLiCationEnvironment2006, borodinMechanismIonTransport2006, liEffectOrganicSolvents2015, leschInfluenceCationsLithium2016}
This first coordination shell is fully contained within a distance of $r_\text{fc} = \SI{3}{\angstrom}$ from the cation.
Based on this observation all oxygen atoms with a distance $r \leq r_\text{fc}$ can be defined as coordinating a cation.

\begin{figure*}
 \centering
 \includegraphics{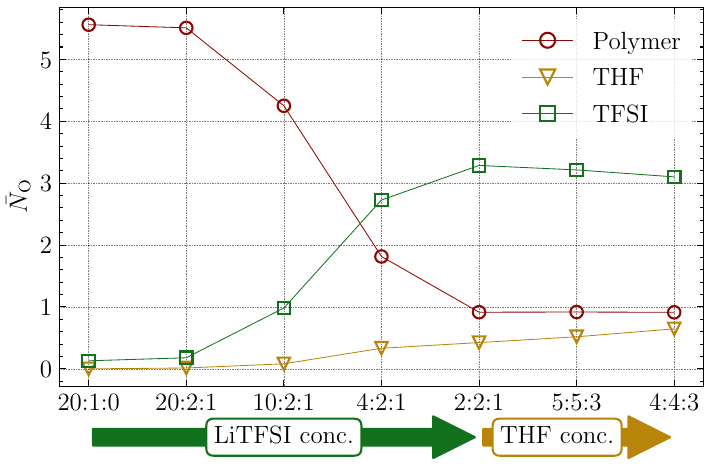}
 \caption{\label{fig:coord}Average number $\bar{N}_{\ce{O}}$ of cation--coordinating oxygen atoms for the various ligand residues.}
\end{figure*}

By counting the number of oxygen atoms within the first coordination shell of the cations the coordination number can be determined.
The average coordination number $\bar{N}_{\ce{O}}$ for oxygen atoms of the various residues in the simulated systems is shown in \cref{fig:coord}.

In systems with a \ce{LiTFSI} concentration of \sysFive{} and lower, the average coordination number of polymer oxygen is around five and thus matches other reported values.
\cite{muller-platheComputerSimulationPolymer1995, borodinMolecularDynamicsSimulations1998, borodinMechanismIonTransport2006}
Generally the number of coordinating polymer oxygens decreases with increasing salt concentration for all simulated systems.
The largest reduction of $\bar{N}_{\ce{O}}$ for the polymer occurs when increasing the salt concentration from \sysFive{} to \sysTwo{}.
The resulting coordination number in the \sysTwo{} system is just below two.
It can be argued that the \sysFive{} system provides just enough oxygen atoms from the polymer so that the PEO chains can wrap around the cations.
For higher salt concentrations the PEO chains might unwrap and stretch to accommodate more cations on the chain which explains the lower number of coordinating polymer oxygens in the \sysTwo{} system.
In the \sysTwo{} system the polymer chains are saturated with cations and consequently the number of cations per polymer chain cannot increase further when adding salt (see SI \cref{fig:si_coord_res_li} (left)).
The proposed unwrapping of the PEO chains allows for other ligands to coordinate cations too which leads to mixed coordination environments as discussed next.

Contrasting with the coordination number of polymer oxygens, the number of coordinating \ce{TFSI} oxygens generally increases for higher salt concentrations.
In systems with a salt concentration of \sysFive{} and lower, $\bar{N}_{\ce{O}}$ for \ce{TFSI} is below one.
In the \sysTwo{} and \sysOneGroup{} systems, $\bar{N}_{\ce{O}}$ for the anions is around three.
Therefore, the largest increase of the number of coordinating \ce{TFSI} oxygen atoms can be observed between the \sysFive{} and \sysTwo{} systems.
Again this can be explained by the saturation of the polymer chain with coordinating cations.
PEO oxygens are preferred for coordinating cations but, if there are not enough free polymer oxygens available, \ce{TFSI} coordinates instead.
Furthermore, the short tempering run at $T = \SI{450}{\kelvin}$, which is part of the equilibration procedure,
effectively decreases the number of polymer oxygens while increasing the number of \ce{TFSI} oxygens in cases of high salt loadings (cf.~SI \cref{fig:si_nonequi_coord_pol_tfsi_ts}).
Between two and three coordinating cations per anion can be observed in the \sysTwo{} and \sysOneGroup{} systems (cf.~SI \cref{fig:si_coord_res_li} (left)).
A single \ce{TFSI} molecule is typically able to coordinate a \ce{Li+} with either one (monodentate) or two oxygen atoms (bidentate).
\cref{fig:si_coord_res_li} (right) in SI shows that the majority of cations is only coordinated by monodentate \ce{TFSI} in the \sysFive{} system and systems with less salt.
This fraction steeply increases with the addition of salt between the \sysTen{} and \sysFive{} systems
and reaches a plateau of $p \approx \SI{42}{\percent}$ for all higher salt concentrations.
The previously observed increase of the coordinating \ce{TFSI} oxygen atoms between the \sysFive{} and \sysTwo{} systems can be attributed to an increase of bidentate \ce{TFSI}.
In the \sysTwo{} and \sysOneGroup{} systems the majority of cations is coordinated by at least one bidentate \ce{TFSI}.

The average coordination number of \ce{THF} is below one in all systems and it is the smallest of all three ligand residues.
For the systems with a salt concentration of \sysFive{} and lower, virtually no coordination of the cations by \ce{THF} can be found.
In the \sysTwo{} and \sysOneGroup{} systems $\bar{N}_{\ce{O}}$ for \ce{THF} increases slightly which again can be explained by the saturation of the polymer chain with cations.
Similar to the anions \ce{THF} substitutes the missing PEO oxygen in the cation coordination.
This substitution however is not as pronounced as for the \ce{TFSI}.
When the \ce{THF} concentration is increased in the \sysOneGroup{} systems $\bar{N}_{\ce{O}}$ for \ce{THF} increases as well,
while $\bar{N}_{\ce{O}}$ for the \ce{TFSI} coordination decreases slightly.
In those systems \ce{THF} seems to be able to replace some coordinating anion oxygens.
The coordination number of polymer oxygens does not change upon increased \ce{THF} content.

\subsection{Number of Coordinating Residues}
\begin{figure}
 \centering
 \includegraphics{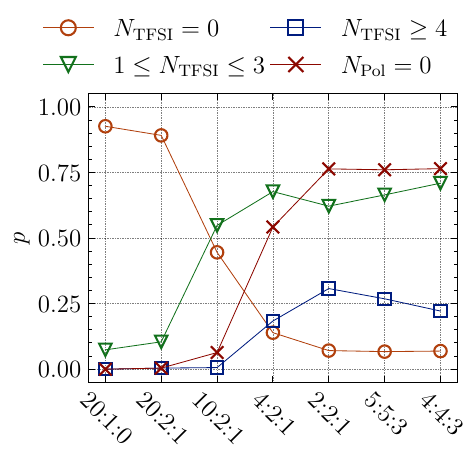}
 \caption{\label{fig:coord_mol}Fraction $p$ of cations with a certain number of coordinating residues.}
\end{figure}

Next we investigate the \ce{Li+} coordination in more detail by analyzing the number of coordinating residues.
A residue is counted as coordinating if at least one of its oxygen atoms resides within the first coordination shell of a given cation.
In \cref{fig:coord_mol} the fraction $p$ of cations with a certain number of coordinating residues is shown.

Systems with a salt concentration of \sysFive{} and lower have nearly all of their cations coordinated by the polymer which is indicated by a low fraction of $N_{\text{Pol}} = 0$.
Moreover, in systems with a salt concentration of \sysTen{} and lower, most cations are exclusively coordinated by polymer, reflected by a high fraction of $N_{\ce{TFSI}} = 0$.
Increasing the \ce{LiTFSI} concentration from \sysFive{} to \sysTwo{} results in a sharp rise of the fraction of cations with $N_{\text{Pol}} = 0$.
In the \sysTwo{} system less than half of all cations are coordinated by PEO chains.
The fraction of cations that are not coordinated by polymer further rises to around \SI{75}{\percent} in the \sysOneGroup{} systems.
As before, the addition of \ce{THF} in those systems does not change the polymer coordination.
Further investigations reveal that cations coordinated by the polymer have typically only one attached PEO chain.
Only in the \sysTwenty{} and \sysTen{} systems cations with two coordinating PEO chains are found in low numbers ($p < \SI{2.5}{\percent}$).
The low occurrence of such coordination environments has already been discussed in reference \citen{diddensImprovingLithiumIon2017}.

The inverse behavior between the fraction of cations that have no anion coordination ($N_{\ce{TFSI}} = 0$) and cations without polymer coordination ($N_{\text{Pol}} = 0$) is as expected.
The fraction of cations with $N_{\ce{TFSI}} = 0$ generally declines for higher salt concentrations.
In the \sysTwenty{} and \sysTen{} systems most cations do not have coordinating anions,
whereas more than half of all cations are coordinated by one to three \ce{TFSI} in the \sysFive{} system.
Higher coordination numbers are not observed in this system.
In the \sysTwo{} system most of the cations are coordinated by up to three anions.
The fraction of cations with $N_{\ce{TFSI}} = 0$ further decreases to approximately \SI{15}{\percent}.
Notably roughly the same number of cations have four or more coordinating \ce{TFSI} ($N_{\ce{TFSI}} \geq 4$).
By increasing the salt concentration to \sysOne{} the fraction of cations with $N_{\ce{TFSI}} \geq 4$ increases while the fraction of cations without \ce{TFSI} decreases.
These high coordination numbers are especially prevalent in the salt--rich layer in the middle of the bilayer as discussed later on.
The fraction of cations that have one to three neighboring anions declines slightly due to the overall higher number of \ce{Li+} and the pronounced central salt--rich layer.

\begin{figure*}
 \centering
 \includegraphics[width=0.8\textwidth]{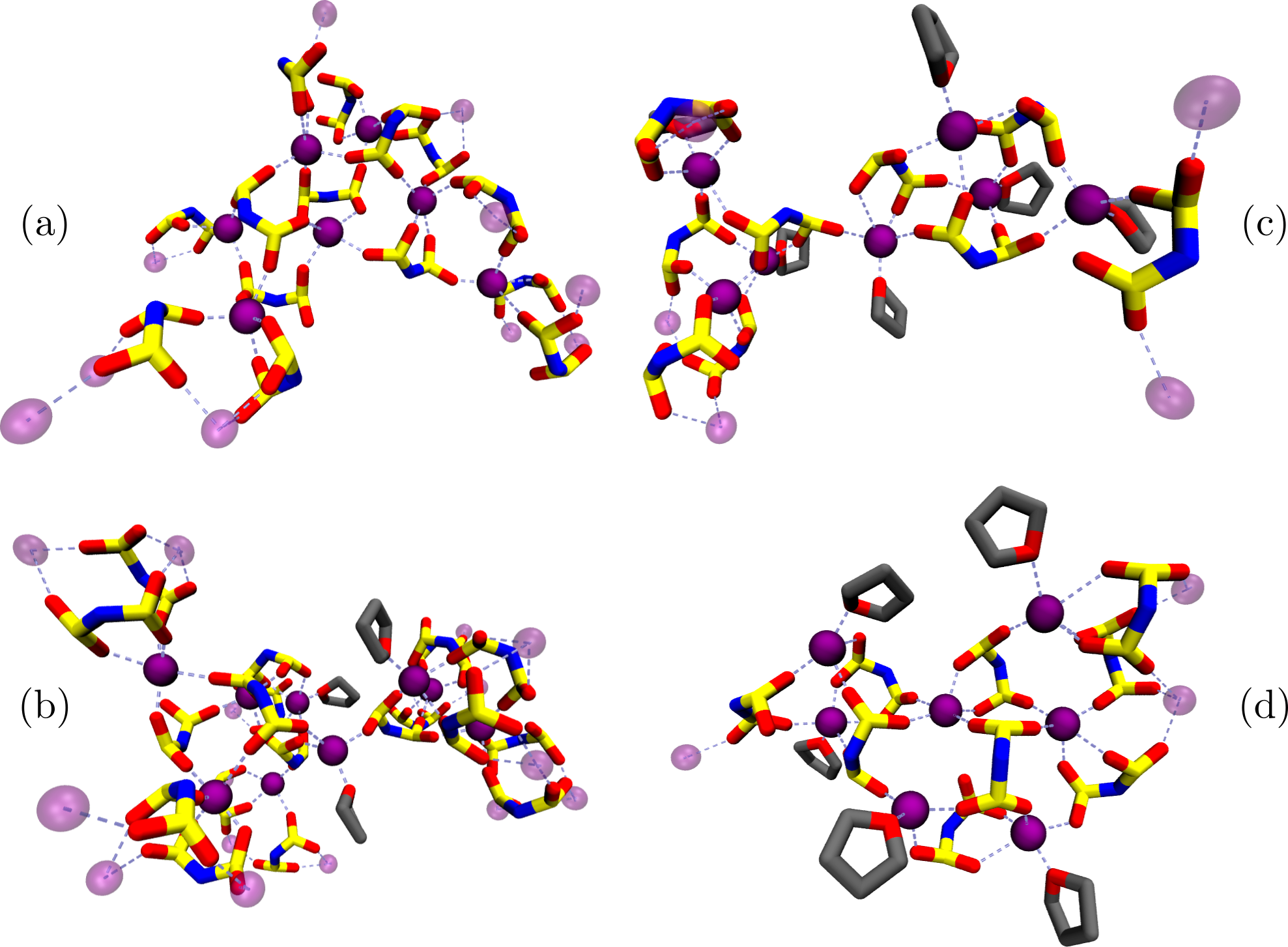}
 \caption{\label{fig:snapshot_clusters}Snapshots of common substructures in the large cluster formed in the central salt--rich layer in the \sysOneGroup{} systems.
 Representative motifs were taken from the \sysOne{} system (a, b) and the \sysOneThfB{} system (c, d).
 A substructure contains the molecules of the inner three coordination shells around a central cation.
 The overall network continues further in places indicated by translucent \ce{Li+}.
 The \ce{-CF3} groups of \ce{TFSI} and hydrogen of \ce{THF} are omitted for clarity.
 Oxygen atoms and \ce{Li+} are shown in red and purple respectively.
 Coordinations between \ce{Li+} and \ce{O} are indicated by dashed lines.
 Nitrogen is shown in blue, sulfur in yellow, and carbon in gray.}
\end{figure*}

The previous observations seem to suggest the formation of a network--like structure constructed from anions that are linked by cations in systems with a salt concentration of \sysTwo{} and above.
This is because anions coordinate to multiple cations and vice versa.
The proposed network--like structure is characterized in detail by determining the sizes of clusters formed by anion--cation aggregates (cf.~SI \cref{fig:si_clusters}).
The cluster size is defined as the number of cations interconnected by bridging anions.
Cations bridged by PEO chains are not considered to be in the same cluster.
In the \sysTwenty{} and \sysTen{} systems virtually no anion--cation aggregates are formed matching the observed low coordination numbers of \ce{TFSI} around \ce{Li+}.
In the \sysFive{} system approximately \SI{30}{\percent} of the cations are part of small clusters consisting of less than 10 cations.
Contrary to the low salt systems most cations are part of a single large network in the \sysTwo{} and \sysOneGroup{} systems.
In the \sysTwo{} system just slightly more than half of all cations are part of this network.
But in the \sysOneGroup{} systems more than \SI{80}{\percent} of cations are interconnected by \ce{TFSI}.
Comparable observations have also been made in concentrated mixtures with other salts and solvents,
and it has been proposed that such structures favor a hopping based transport mechanism.\cite{seoElectrolyteSolvationIonic2013, dokkoDirectEvidenceLi2018}
\Cref{fig:snapshot_clusters} shows common structural motifs found in this network.
Additional \ce{THF} in those systems detaches \ce{TFSI} from highly coordinated cations and is therefore able to partially break up the network--like structure
as indicated by a reduction of $N_{\ce{TFSI}} \geq 4$ in \cref{fig:coord_mol}.
This results in a reduced fraction of cations contained within the network (cf.~SI \cref{fig:si_clusters}).
The partial break up of the network is also visually supported by the observation that \ce{THF} seems to be able to separate highly interconnected regions (cf.~\cref{fig:snapshot_clusters}).
The atomistic insights presented here corroborate hypothesized structures from experimental studies.\cite{zhou_deciphering_2021,krause_superionic_2024}

To understand why the polymer coordination drops steeply between the \sysFive{} and \sysTwo{} systems the coordination number is spatially resolved across the bilayer (cf.~SI \cref{fig:si_coord_profile}).
For systems with a salt concentration of \sysFive{} and lower the coordination of PEO, \ce{TFSI} and \ce{THF} is homogeneous across the PEO domain.
This holds true for the number of coordinating atoms as well as residues (not shown).
The \sysTwo{} system exhibits two differences when compared to systems with lower salt concentrations:
First there is a significant amount of cations between the PEO chains that are not directly attached to those chains (see $N_{\ce{O}} = 0$ at $z/L_0 = 0.4$ and $0.6$).
They are coordinated by anions and \ce{THF} only and constitute approximately \SI{50}{\percent} of the cations that reside within the PEO domain
(note that the overall fraction of cations without neighboring PEO in \cref{fig:coord_mol} is higher because of the additional cations in the salt layer).
Consequently, the PEO chains have to be further apart to accommodate the salt.
This is also reflected by the $L_{1,2}$ values in \cref{tab:box_dims} which increase with higher salt concentrations.
Second the distance between the top and bottom layer of the block copolymer bilayer increases which causes diminished coordination by PEO in the middle of the bilayer.
Presumably the salt cannot be fully accommodated by lateral swelling of the PEO domain and a salt--rich layer in the middle of the bilayer is formed.
As discussed in context of \cref{fig:res_profile,fig:snapshot_3to2to1}, the swelling increases further in the \sysOneGroup{} systems so that the PEO domain becomes completely separated by a layer in the middle where no polymer coordination of \ce{Li+} is observed.
This assumption is in accordance with the experimental SAXS data in references \citen{dorrAmbientTemperatureElectrolyte2018, pelzSelfAssembledBlockCopolymer2019}.
In this layer the cations are coordinated predominantly by \ce{TFSI} but upon adding \ce{THF} in the \sysOneGroup{} systems this additive increasingly takes part in the coordination as well.

\subsection{Polymer Structure}
\begin{figure*}
 \centering
 \includegraphics{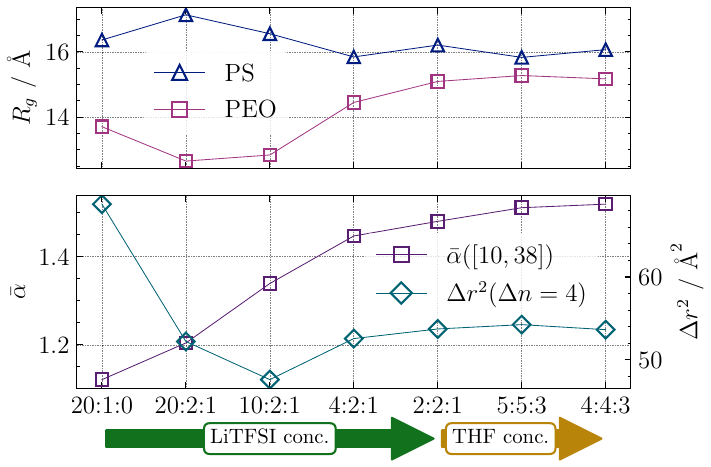}
 \caption{\label{fig:pol_struct}Radius of gyration $R_g$ for the PS and PEO block (top).
 Estimated long--scale stretching parameter $\hat{\alpha}$ and short--scale squared distance $\Delta r^2$ for the PEO block (bottom).}
\end{figure*}

As a first assessment of the structure of the block copolymers the radius of gyration $R_g$ is calculated for the PS and PEO block independently.
The resulting values of $R_g$ are shown in \cref{fig:pol_struct} (top).

The radius of gyration for the PS chains generally decreases if the \ce{LiTFSI} concentration is increased.
This is not the case for the \sysTwenty{} system because $R_g$ in this system is lower than $R_g$ in the next higher concentrated system.
Presumably this might be related to the fact that the \sysTwenty{} system does not contain any \ce{THF}.
Therefore, one can reasonably assume that \ce{THF} as a solvent within the PS block allows stretching of the PS chains in the \sysTen{} system.\cite{milnerPolymerBrushes1991}
The previously discussed mass density within the bilayer (cf.~\cref{fig:res_profile}) shows however that with an increased number of \ce{THF} molecules the density of \ce{THF} within the PS block expectedly increases as well.
At first glance this contradicts the observed decrease in the radius of gyration for higher \ce{LiTFSI} concentrations because it is expected that \ce{THF} stretches the chains.
As discussed in the previous section, the lateral size of the simulation box $L_{1,2}$ increases upon adding salt.
Consequently, the PS chains have to be further apart so that a more compact conformation is adopted to fill the created interstice which results in a smaller $R_g$.
Complementary one can argue that the increasing amount of salt within the PEO block immobilizes the block copolymers
and prompts a clearly defined phase separation between apolar PS chains and PEO embedded in salt.\cite{chuIonDistributionMicrophaseSeparated2018}
Analysis of the mean squared displacement (MSD) of both the EO and PS monomers display decreasing mobility for systems with more salt (not shown).
Additionally, the ability of salt to promote demixing is also reported in references \citen{youngSaltDopingPEOContaining2009, nakamuraSaltdopedBlockCopolymers2012, teranThermodynamicsBlockCopolymers2014, qinOrderingTransitionSaltDoped2016}.
Both effects may lead to compression of the PS block and $R_g$ for the PS chains decreases.
This compression effect seems to be limited because the observed change between the \sysTwo{} and \sysOneGroup{} systems is negligible.
The systems with a salt concentration of \sysOneGroup{} exhibit no pronounced influence of additional \ce{THF} because the added solvent predominantly resides within the polar salt layer.

The radius of gyration for the PEO chains is minimal for a salt concentration of \sysTen{} and \sysFive{}.
The PEO chains exhibit the least amount of stretching in those systems because most cations are coordinated by PEO in such a way that the PEO chains are wrapped around the cations.
Fewer cations that cause wrapping of the PEO chains result therefore in a higher $R_g$ in the \sysTwenty{} system.
Salt concentrations of \sysTwo{} and higher result in a higher $R_g$ as well because the PEO chains are saturated with cations and have to stretch to accommodate additional cations.
Similar behavior is also captured by the end--to--end distance of the PEO chains which complements $R_g$ (cf.~SI \cref{fig:si_ree}).

To investigate the stretching of the PEO chains further the squared distance $\Delta r^2$ between two EO monomers of the same chain can be described
as a function of the number $\Delta n$ of interjacent monomer--monomer bonds.
It is expected that this distance generally follows the form $\Delta r^2(\Delta n) \propto \Delta n^\alpha$.
The parameter $\alpha$ can be called stretching parameter.
In case of a fully elongated chain the stretching parameter has a value of $\alpha = 2$.
It can be shown that if the chain describes a random walk the stretching parameter has a value of $\alpha = 1$.\cite{guthElasticThermoelasticProperties1941}
Since the block copolymers assemble in a bilayer, the PEO chains are not fully free to move.
They are fixed in place by cohesion of the PS blocks.
Milner \cite{milnerPolymerBrushes1991} proposes a model which describes the elongation of polymer chains that form such a brush--like structure.
This approach describes fully elongated chains ($\alpha = 2$) in the presence of an ideal solvent and less stretched chains with $\alpha \approx 1.3$ without a solvent.

It is possible to compute a local estimate for the stretching parameter of a segment with length $\Delta n$ as numerical derivative of $\log\left( \Delta r^2 \right) \propto \alpha \log\left( \Delta n \right)$:
\begin{equation}\label{eq:alpha_local}
 \hat{\alpha}(\Delta n) = \frac{\log\left( \frac{\Delta r^2(\Delta n - 1)}{\Delta r^2(\Delta n + 1)} \right)}{\log\left( \frac{\Delta n - 1}{\Delta n + 1} \right)}.
\end{equation}
Numerical analysis of \eqref{eq:alpha_local} reveals three distinct aspects for segments of length $\Delta n$ (cf.~SI \cref{fig:si_alpha_local}):
Short segments exhibit a minimum of $\hat{\alpha}$ at $\Delta n = 4$.
A plateau of $\hat{\alpha}$ is reached for medium--sized segments in the interval $\Delta n \in [10, 38]$.
For long segments the approximation of the stretching parameter is influenced by finite--size effects at the ends of the chains.
Those effects arise from the entropic repulsion of distinct segments due to excluded volume effects, which is less pronounced for the chain ends.\cite{wittmerIntramolecularLongrangeCorrelations2007}
As argued in the previous sections the PEO chains wrap around cations with up to five continuous oxygen atoms, i.e.~segments of lengths $\Delta n = 4$.
Since this exactly coincides with the local minimum at $\Delta n = 4$, one can reasonably assume that the PEO--chain structure is dominated by cation--induced curvature on a length scale with $\Delta n < 10$.
To describe the stretching on this small scale the squared distance $\Delta r^2(\Delta n = 4)$ is used in the following.
This value directly correlates with the minimum value of $\hat{\alpha}(4)$ (cf.~SI \cref{fig:si_rsq_4}).
The overall chain structure of PEO is described on a larger length scale.
In the following the average
\begin{equation}
 \bar{\alpha}([10, 38]) = \frac{1}{38 - 10 + 1} \sum_{\Delta n = 10}^{38} \hat{\alpha}(\Delta n)
\end{equation}
of the plateau values is used to describe the stretching on this larger scale.
The long--scale stretching parameter $\bar{\alpha}([10, 38])$ and the short--scale squared distance $\Delta r^2(\Delta n = 4)$ are shown in \cref{fig:pol_struct} (bottom) for all simulated systems.

The squared distance $\Delta r^2(\Delta n = 4)$ exhibits a minimum for the \sysFive{} system.
As before one can argue that for this concentration of salt all monomers of the PEO chain are involved in wrapping coordinated cations.
If there are fewer cations available, not all parts of a PEO chain are compressed by being wrapped around cations.
As a consequence, $\Delta r^2(\Delta n = 4)$ increases with lower salt concentrations.
If the salt concentration is higher than \sysFive{}, the PEO chain is stretched to accommodate more cations.
Additional \ce{THF} has no influence on $\Delta r^2$ for short length scales in the \sysOneGroup{} systems.

The stretching parameter $\bar{\alpha}([10, 38])$ for the overall chain increases with higher salt concentrations.
This suggests that the concentrated \ce{LiTFSI}/\ce{THF} electrolyte can be interpreted as solvent that favorably interacts with the PEO chains.
When considering the reduced number of coordinating PEO oxygens and small number of \ce{Li+} that are exclusively bound to the polymer in salt--rich systems,
it becomes apparent that this interaction originates from cations that are partially bound to PEO and also part of \ce{LiTFSI} clusters.
In order to maximize the interaction with the salt the PEO chains are stretched.
This might also mitigate the electrostatic repulsion between two neighboring cations that are bound to the same PEO chain.
However, a salt--rich layer nonetheless forms once the PEO chains become saturated.
Even in the systems with a high salt content fully elongated chains ($\alpha = 2$) are never reached due to entropical reasons.
Also in the system with the lowest amount of salt the chains are less stretched than expected from the brush model without an ideal solvent ($\alpha \approx 1.3$).
One can reason that the stretching parameters are lower than expected from the brush model for both low and high salt concentrations because the cations induce local curvature and thus cause less stretching.
In addition, the chains in the simulation are shorter and may therefore be more stretched.
But even under low salt conditions the chains are still far from describing a random walk ($\alpha = 1$) due to excluded volume effects in densely packed brushes.\cite{milnerPolymerBrushes1991}
In the \sysOneGroup{} systems no clear influence of additional \ce{THF} can be observed.

\subsection{\label{sec:cond}Cationic Conductivity}
The MSD $\langle \Delta r_i^2(\Delta t) \rangle$ can be determined for an ionic species $i$
and subsequently allows the calculation of the diffusion coefficient $D_i$ by utilizing the Einstein relation:\cite{borodinLiTFSIStructureTransport2006}
\begin{equation}\label{eq:einstein}
 D_i = \lim_{\Delta t \rightarrow \infty} \frac{\langle \Delta r_i^2(\Delta t) \rangle}{6 \Delta t} .
\end{equation}
The displacement during a time $\Delta t$ is calculated for \ce{Li+} and the nitrogen of \ce{TFSI} (cf.~SI \cref{fig:si_msd_li,fig:si_msd_tfsi}).
$\langle \dots \rangle$ denotes the ensemble average.
The calculated diffusion coefficients are listed in SI \cref{tab:si_diff_ions}.
Both the diffusion coefficient of the cations and of the anions become generally smaller for an increase of salt,
but the motion of \ce{Li+} becomes diffusive after a shorter period of time which indicates a decoupling from the PEO motion.
In the systems with a lower salt concentration the anions are much more mobile, but in the \sysOneGroup{} systems the diffusion coefficients of anions and cations are approximately equal.
Those two observations have also been made previously in concentrated liquid electrolytes.\cite{selfTransportSuperconcentratedLiPF2019, alvaradoCarbonatefreeSulfonebasedElectrolyte2018, dongHowEfficientLi2018}
The results of this study regarding the coordination numbers suggest that anions and cations form a network--like structure for high salt concentrations.
Anions are slowed down relative to the \ce{Li+} motion by coordinating to multiple cations.
Remarkably the \sysOneGroup{} systems with additional \ce{THF} exhibit higher cationic diffusion coefficients than the \sysOne{} and \sysTwo{} systems which has also been observed experimentally.\cite{krause_superionic_2024}
When compared to the \sysOne{} system \ce{THF} increases the ion mobility roughly by a factor of two in the \sysOneThfA{} system and five in the \sysOneThfB{} system (cf.~SI \cref{tab:si_diff_ions}),
although only \SI{20}{\percent} and \SI{50}{\percent} of the minority component \ce{THF} are added respectively in those systems.

The cationic transport mechanism in relation to \ce{THF} is further explored by a set of simulations without polymer which mimic the salt--rich central layer of the polymer systems by composition.
A wider range of \ce{THF} concentrations is explored in those simulations as well (see \cref{tab:si_systems_salt_only} in SI).
The time--dependent cationic MSD is observed for cations which are coordinated by \ce{THF} at the starting time and is given relative to the displacement of cations with no \ce{THF} in their first coordination shell.
\Cref{fig:si_msd_li_thf_salt} shows that generally \ce{Li+} with \ce{THF} has a larger displacement than cations without \ce{THF}.
On long time scales the displacement of both groups becomes equal because the coordination environment undergoes exchange with the surroundings and loses similarity to the starting condition.
This exchange happens faster for higher amounts of \ce{THF} but the decay is in the order of \SI{100}{\nano\second} in all simulated systems.
At the same time, the relative improvement of the transport is largest for high amounts of \ce{THF}.
The relative improvement happening on long time--scales shows that \ce{THF} not only accelerates the overall transport by reducing the viscosity
but also improves the transport locally.
Considering the structural motifs observed in \cref{fig:snapshot_clusters},
this accelerated transport can be justified by the ability of \ce{THF} to break up \ce{LiTFSI} clusters and promote \ce{Li+} hopping between anions.

\begin{figure*}
 \centering
 \includegraphics{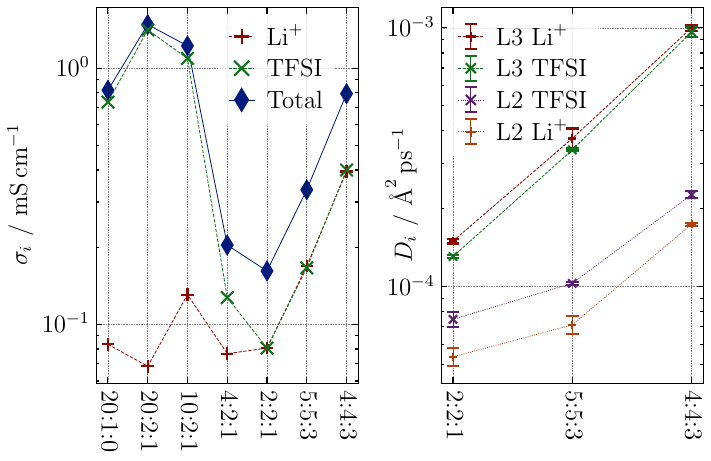}
 \caption{\label{fig:ideal_cond}Nernst--Einstein conductivity $\sigma_i$ for \ce{Li+}, \ce{TFSI}, and both species combined (left).
 Lateral diffusion coefficients $D_i$ in layers 2 and 3 for ions in the \sysOneGroup{} systems (right).}
\end{figure*}

As a proxy for the overall ionic conductivity the Nernst--Einstein conductivity $\sigma_i$ of individual species can be calculated as follows:\cite{muller-platheComputerSimulationPolymer1995}
\begin{equation}
 \sigma_i = \frac{e^2}{V k_B T} \cdot N_{i} D_{i} .
\end{equation}
In this equation $e$ denotes the electron charge, $V$ the average volume of the simulation box, $T$ the temperature, and $N_i$ the number of ions per system.
The subscript $i$ denotes \ce{Li+} and \ce{TFSI} respectively and the total conductivity is obtained by summing over the contributions of both ion species.
The resulting values of the Nernst--Einstein conductivity and are shown in \cref{fig:ideal_cond} (left).
However, these simplified measures capture transport due to self-diffusion only and are unable to consider ion correlations.\cite{borodinLiTFSIStructureTransport2006, borodinLiTransportMechanism2007}
The calculation of the overall ionic conductivity including ion correlations necessitates a longer simulation time to allow better statistical analysis.
Nonetheless, a promising increase of the Nernst--Einstein conductivities for cations can be observed for systems with a high salt loading and additional \ce{THF}.
The \sysOneGroup{} systems additionally exhibit approximately equal contributions of anions and cations to the conductivity
which results in an increased \ce{Li+} transference number when compared to systems with less salt.
A superior lithium transference number has also been attested experimentally.\cite{pelzSelfAssembledBlockCopolymer2019}
However, a simple quantitative comparison between these conductivities and the corresponding experimental values is not possible because of differing conditions:
The higher temperatures in the simulations should increase the simulated conductivities in comparison to the experiment.
Contrary the much thicker salt layer between the PEO domains in the experiment would increase the experimental values.

In order to get a better understanding of the nature of the cation transport the lateral $\text{MSD}_{xy}$ was calculated in three distinct layers within the bilayer (cf.~\cref{fig:si_msd_layer}):
Layer 1 contains the interface between PS and PEO domains.
Layer 2 hosts the bulk of the PEO and layer 3 contains the PEO--chain ends as well as the central salt layer.
The ions are assigned to the appropriate layer based on their $z$ position at beginning of the time interval $\Delta t$.
This observation reveals that the displacements in layer 1 and 2 are comparable to each other.
However, the displacement in layer 3 is much larger than in the other layers.
The slower movement of the cations between the polymer might be caused by the slower movement and higher viscosity of the polymer itself in addition to the strong coordination between \ce{Li+} and PEO.

Furthermore, the lateral diffusion coefficients in the layers can be calculated for both ion species based on the lateral $\text{MSD}_{xy}$.
Since layer 1 contains fewer ions than the other two layers the diffusion coefficient in this layer is more susceptible for statistical fluctuations.
Consequently, the diffusion coefficients are calculated only for layers 2 and 3 which accommodate most of the ions.
The resulting values for the \sysOneGroup{} systems are shown in \cref{fig:ideal_cond} (right) and reflect the observations made above (see also SI \cref{tab:si_diff_ions_layer}).
In all \sysOneGroup{} systems a maximal diffusion coefficient is observed for both ion species in layer 3 and the increase caused by additional \ce{THF} is more pronounced in this central layer.
But, most importantly, there is another difference between the layer that contains most of the polymer chains (i.e.~layer 2) and the central salt--rich layer (i.e.~layer 3):
In layer 2 the anionic diffusion coefficient is always larger than the cationic diffusion coefficient.
Contrary in layer 3 the cationic diffusion coefficient is equal to or larger than the anionic diffusion coefficient in all \sysOneGroup{} systems.
This again can be attributed to the strong coordination between PEO chains and cations in layer 2 which hinders cationic movement.

As a consequence the salt--rich layer in the middle of the bilayer seems to play an important role in facilitating a high ionic conductivity and a large cationic transference number.
Most likely, this is even more pronounced in the experimental systems which have a much thicker salt layer.
In recent years the concept of ``solvent--in--salt'' electrolytes is being actively explored again
because an unusually high concentration of salt promises improved electrochemical stability of the electrolyte, advantageous electrode interface properties, and high lithium transference numbers.
\cite{suoWaterinsaltElectrolyteEnables2015, qianHighRateStable2015, chenElucidationTransportMechanism2016, borodinUnchartedWatersSuperConcentrated2020}
Super--concentrated block copolymer systems as discussed in this work might provide a way to combine those electrochemical advantages with additional mechanical rigidity.
These findings motivate future work on the transport in such systems.

\section{Conclusion}
In the present study, we elucidated the structural properties of lamellar BCPs with unusually short PEO chains and high \ce{LiTFSI} salt loadings.
It was found that increasing the salt concentration from \sysFive{} to \sysTwo{} (i.e.~EO:\ce{Li+}:\ce{THF}) results in a change of the cation--coordination environment.
A steep drop of the number of coordinating PEO oxygen atoms and an increase of coordinating \ce{TFSI} oxygens has been observed in this case.
The \sysTwo{} system exhibits a fraction of more than half of all cations that are no longer coordinated by any PEO chain.
Most notably, this change in coordination is not only caused by a swelling of the PEO domain but also by the formation of a salt--rich layer in the middle of the bilayer.
Starting from intermediate salt concentrations, the systems show the retraction of PEO from this central layer.
Consequently, cations are coordinated by \ce{TFSI} and \ce{THF} exclusively in the middle.
The added salt also causes stretching of the PEO chains in order to accommodate more cations per polymer chain.
As a result, the number of \ce{Li+} per polymer chain strongly increases.
A similar increase in coordination is observed with respect to the anions as well.
Since multiple cations coordinate a single anion and vice versa, a network--like structure of salt is formed between the polymer chains.
In the limit of large salt concentrations the majority of cations is part of a single large cluster.
Additional \ce{THF} is able to partially break up this structure by removing \ce{TFSI} from highly coordinated cations.

Most importantly, systems with high salt concentrations exhibit remarkably high cationic mobility located in the central salt--rich layer unraveling the microscopic origin of the experimental observations.
Unlike typical PEO--based SPEs in which the anions show much higher diffusion coefficients, the diffusion coefficients are approximately equal for both anions and cations.
This qualitatively agrees with the high cationic transference numbers found in the experiment.
Since the PEO chains are immobilized by the high salt loadings, we conclude that this cation transport is decoupled from the polymer motion.
This decoupled mode of transport is generally applicable to a wide range of microstructures with continuous conduction paths.
A key contribution of this work is that the cation transport is facilitated by the \ce{LiTFSI} network in the central salt layer instead.
Furthermore, the ion transport can be strongly improved by adding \ce{THF}.
In the future other solvents with better properties, e.g.~higher electrochemical stability, can be investigated as a \ce{THF} substitute.
Thus, the tuning of transport properties can be achieved by optimizing solvents in experiments.

\providecommand{\latin}[1]{#1}
\makeatletter
\providecommand{\doi}
  {\begingroup\let\do\@makeother\dospecials
  \catcode`\{=1 \catcode`\}=2 \doi@aux}
\providecommand{\doi@aux}[1]{\endgroup\texttt{#1}}
\makeatother
\providecommand*\mcitethebibliography{\thebibliography}
\csname @ifundefined\endcsname{endmcitethebibliography}
  {\let\endmcitethebibliography\endthebibliography}{}

\end{document}


\maketitle

\subsection*{Composition of the simulated systems:}
Dörr et al.~(2018) experimentally used BCPs with ultra--small PEO blocks of \SI{2100}{\gram\per\mol}.
This corresponds to 48 EO monomers (using a molecular weight of a monomer of $M_W(\text{EO}) = \SI{44.053}{\gram\per\mol}$).
The best conducting system reportedly has 4.3 \ce{Li+} per EO i.e.~$N_{\ce{LiTFSI}} = 206.4$ per chain.
Additional experimental analysis has shown a content of $9 \pm 3\;\text{wt}\%$ non--removable \ce{THF} after evaporation
and Zhou et al.~(2021) have reported a residual amount of \ce{THF} around $10\;\text{wt}\%$.
For convenience a value of $10\;\text{wt}\%$ \ce{THF} is adopted in the following argumentation.
The BCPs used in the simulation have shorter PS blocks of 48 monomers in order to improve the simulation speed.
The analysis of the ion distribution shows that the PS domain is irrelevant for the structure of the salt--rich domain and the ion transport.
The total molecular weight of those BCPs can be calculated as $M_W(\text{BCP}) = \SI{7113.480}{\gram\per\mol}$ (using a value of $M_W(\text{St}) = \SI{104.152}{\gram\per\mol}$ for PS monomers).
Subsequently the number of \ce{THF} per chain is calculated as follows:
\begin{equation}
 N_{\ce{THF}} = \frac{0.1 \cdot \left( M_W(\text{BCP}) + N_{\ce{TFSI}} \cdot M_W(\ce{LiTFSI}) \right)}{M_W(\ce{THF})} \approx 92.0,
\end{equation}
using a value of $M_W(\ce{THF}) = \SI{72.107}{\gram\per\mol}$ and disregarding the contribution of \ce{THF} to the overall weight.
Therefore in this best conducting system the number of \ce{THF} per polymer chain is roughly approximated half the number of \ce{LiTFSI}.
Consequently this approximation is used to calculate the number for all simulated systems as $N_{\ce{THF}} = \frac{1}{2}N_{\ce{TFSI}}$.

In addition to the systems discussed in the main text (\cref{tab:systems}), two more \sysOneGroup{} systems have been simulated to explore the extremes of the THF estimation.
A comparison of the composition of all systems is shown in \cref{tab:si_systems}.
\begin{table*}
 \caption{\label{tab:si_systems}Composition of the simulated \sysOneGroup{} systems including systems with THF concentration which is outside the experimentally expected range.}
 \begin{tabularx}{\textwidth}{@{}>{\centering\arraybackslash}X>{\centering\arraybackslash}X>{\centering\arraybackslash}X>{\centering\arraybackslash}X>{\centering\arraybackslash}X@{}}
  \toprule
   $\text{EO} : \ce{Li+} : \text{THF}$ & $N_{\text{Polymer}}$ & $N_{\text{LiTFSI}}$ & $N_{\text{THF}}$ & ${N_{\text{LiTFSI}}} / {N_{\text{THF}}}$ \\ \hline
   \sysOneThfL & 50 & 2400 & 600 & 4.00 \\ \hline
   \sysOne & 50 & 2400 & 1200 & 2.00  \\
   \sysOneThfA & 50 & 2400 & 1440 & 1.67 \\
   \sysOneThfB & 50 & 2400 & 1800 & 1.33 \\ \hline
   \sysOneThfH & 50 & 2400 & 2400 & 1.00 \\
  \bottomrule
 \end{tabularx}
\end{table*}

\begin{figure}[H]
 \centering
 \includegraphics[width=0.8\textwidth]{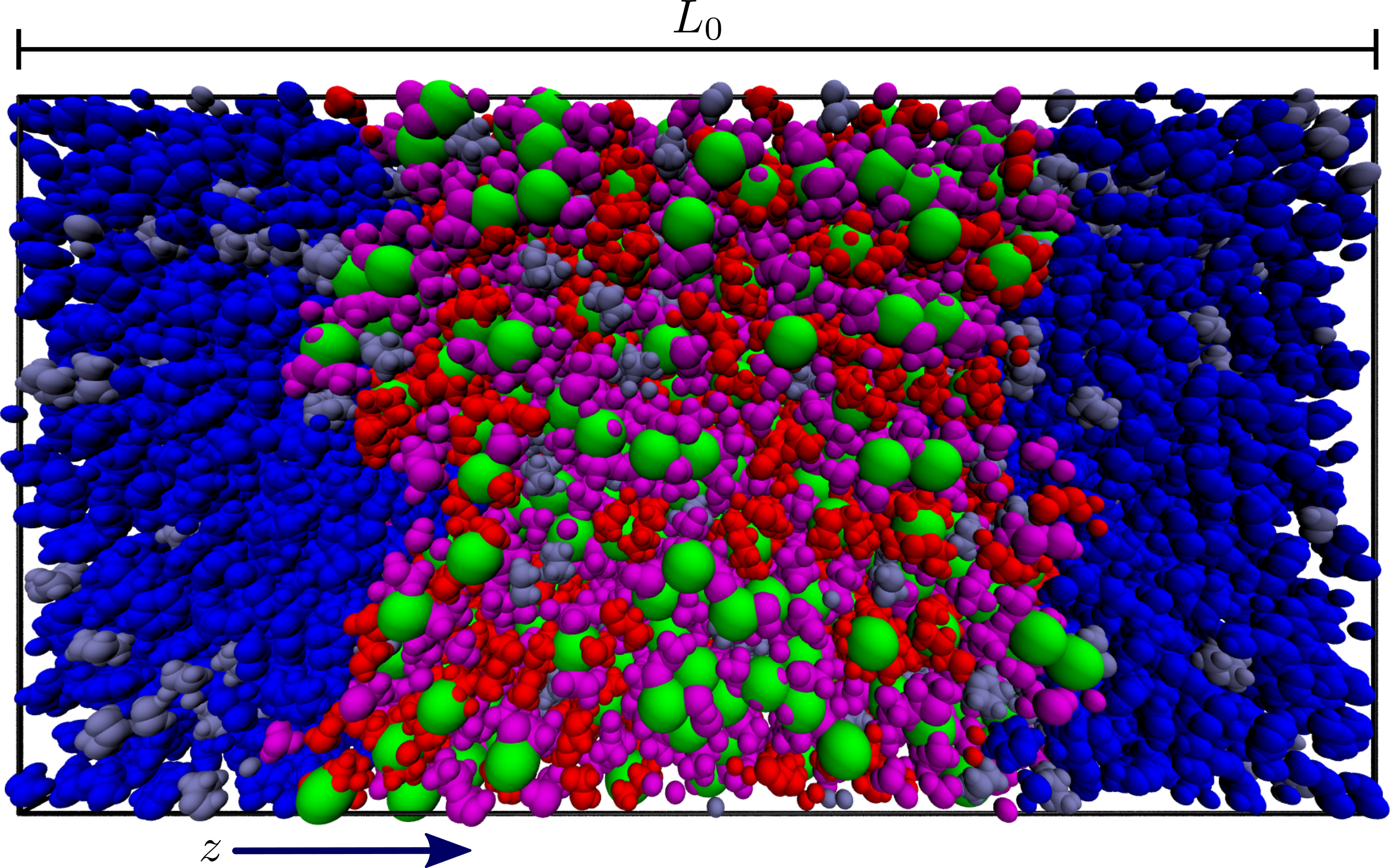}
 \caption{\label{fig:si_snapshot}Snapshot of the \sysTwo{} system with bilayer height $L_0$: PS blue, PEO red, \ce{Li+} green, \ce{TFSI} magenta, \ce{THF} iceblue.}
\end{figure}

\begin{table}[H]
 \caption{\label{tab:box_dims}Bilayer height $L_0$ (which is defined as the height of the simulation box) and lateral dimension $L_{1,2}$ of the simulation box.
 The time average fluctuates with a standard deviation $\sigma$ due to the pressure coupling.}
 \begin{tabularx}{\columnwidth}{@{}>{\centering\arraybackslash}X|>{\centering\arraybackslash}X>{\centering\arraybackslash}X|>{\centering\arraybackslash}X>{\centering\arraybackslash}X@{}}
  \toprule
   System & $L_0$ / \si{\angstrom} & $\sigma(L_0)$ / \si{\angstrom} & $L_{1,2}$ / \si{\angstrom} & $\sigma(L_{1,2})$ / \si{\angstrom} \\ \hline
   \sysTwenty  & 114.190 & 2.035 & 75.849 & 0.682 \\
   \sysTen     & 132.307 & 1.545 & 72.811 & 0.436 \\
   \sysFive    & 140.396 & 1.181 & 74.229 & 0.341 \\
   \sysTwo     & 152.923 & 1.448 & 80.965 & 0.385 \\ \hline
   \sysOne     & 219.702 & 0.845 & 80.120 & 0.156 \\
   \sysOneThfA & 223.423 & 1.088 & 80.448 & 0.202 \\
   \sysOneThfB & 232.452 & 1.719 & 80.317 & 0.300 \\
  \bottomrule
 \end{tabularx}
\end{table}

\begin{figure}[H]
 \centering
 \includegraphics[width=0.8\textwidth]{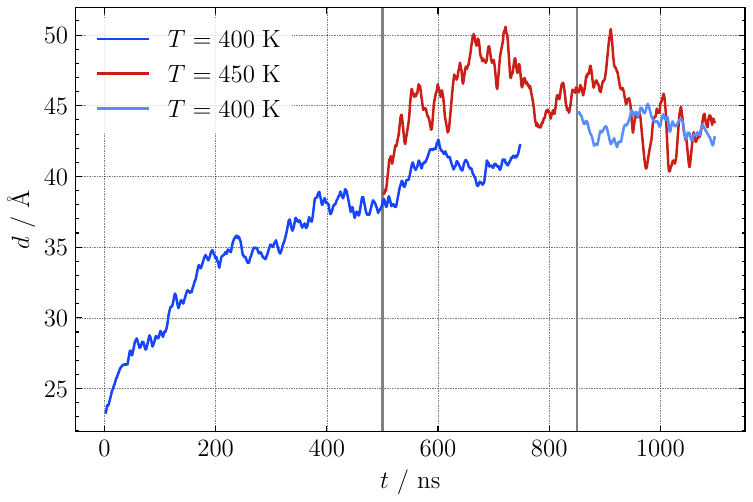}
 \caption{\label{fig:si_nonequi_end_dist_middle_ts}Time series of the distance $d$ between the ends of the PEO chains of the two halves of the bilayer in the \sysEqui{} system.
 An increasing distance signifies the retraction of the PEO chains and formation of a salt--rich layer in the middle of the bilayer.
 A parallel equilibration run at $T = \SI{450}{\kelvin}$ is branching off from the equilibration run at $T = \SI{400}{\kelvin}$ after \SI{500}{\nano\second} to explore higher temperatures.
 Again, after \SI{350}{\nano\second} at a higher temperature a parallel run is branching off which switches back to the lower temperature.
 The observed layer thickness does not change significantly with higher temperatures.
 However, the dimensions of the simulation box irreversibly change after tempering the system as shown in \cref{fig:si_nonequi_box_dims_ts}.}
\end{figure}

\begin{figure}[H]
 \centering
 \includegraphics[width=0.8\textwidth]{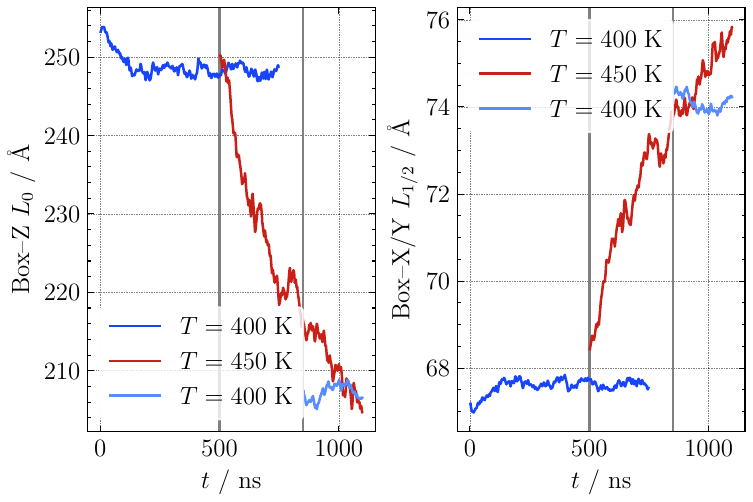}
 \caption{\label{fig:si_nonequi_box_dims_ts}Time series of the dimensions of the simulation box in the \sysEqui{} system.
 Although the bilayer height $L_0$ and the lateral dimension $L_{1,2}$ of the box reach a stable plateau at $T = \SI{400}{\kelvin}$,
 both values are irreversibly changed after tempering at $T = \SI{450}{\kelvin}$.
 It is reasonable to assume that this is dominantly the effect of structural changes in the PS block
 because the PEO domain and the salt layer are not strongly altered (cf.~\cref{fig:si_nonequi_end_dist_middle_ts} and \cref{fig:si_nonequi_coord_pol_tfsi_ts}).}
\end{figure}

\begin{figure}[H]
 \centering
 \includegraphics[width=0.8\textwidth]{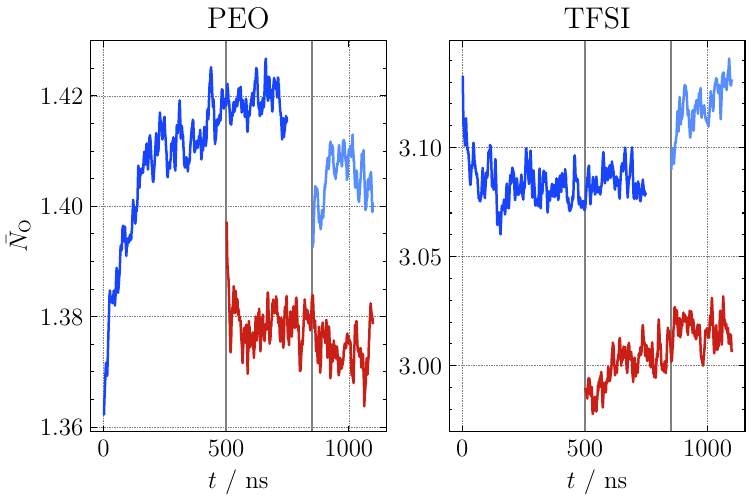}
 \caption{\label{fig:si_nonequi_coord_pol_tfsi_ts}Time series of the average number $\bar{N}_{\ce{O}}$ of coordinating oxygen around \ce{Li+} in the \sysEqui{} system.
 Initially, the average number of PEO oxygen increases hinting at the slow dynamics of the polymer to form local structures around \ce{Li+}.
 The changes caused by tempering are mostly reversible although the average number of coordinating PEO oxygen is reduced
 and the average number of TFSI oxygen is increased after the run at $T = \SI{450}{\kelvin}$.
 This indicates further separation between the PEO domain and the salt--rich layer.
 The cutoff is \SI{3}{\angstrom} which includes the first coordination shell at $T = \SI{400}{\kelvin}$.
 The reduction of the average coordination number at $T = \SI{450}{\kelvin}$ is a combination of less coordination in general
 and a broader distribution of the first shell outside the cutoff radius that was derived from a lower temperature of $T = \SI{400}{\kelvin}$.}
\end{figure}

\begin{figure}[H]
 \centering
 \includegraphics[width=0.8\textwidth]{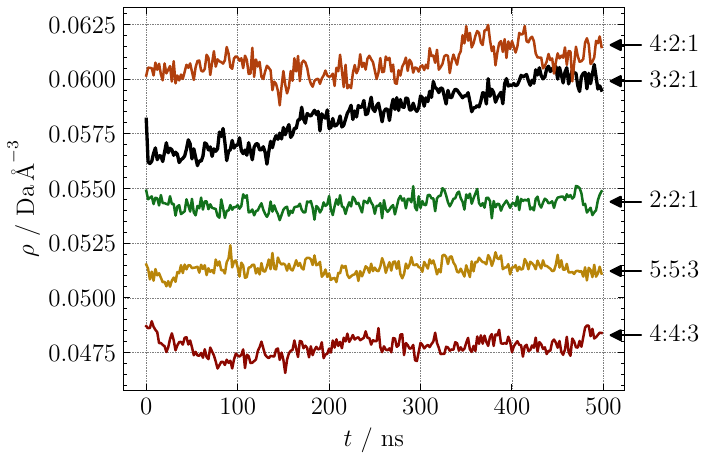}
 \caption{\label{fig:si_peo_profile_ts}Time series of the PEO mass density $\rho$ within the PEO domain ($z/L_0 \in (0.2, 0.4]$) in systems with a \sysTwo{} and higher salt concentration.}
\end{figure}

\begin{figure}[H]
 \centering
 \includegraphics[width=0.8\textwidth]{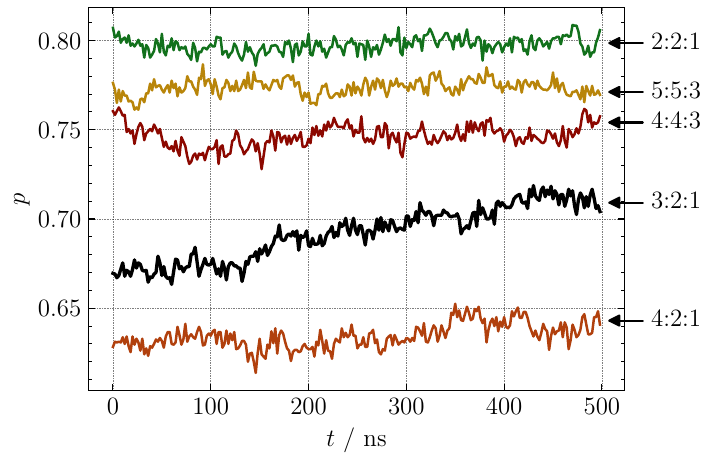}
 \caption{\label{fig:si_peo_p_ts}Time series of the ratio $p$ of EO monomers within the region ($z/L_0 \in (0.2, 0.4]$) in systems with a \sysTwo{} and higher salt concentration.}
\end{figure}

\begin{figure}[H]
 \centering
 \includegraphics[width=0.8\textwidth]{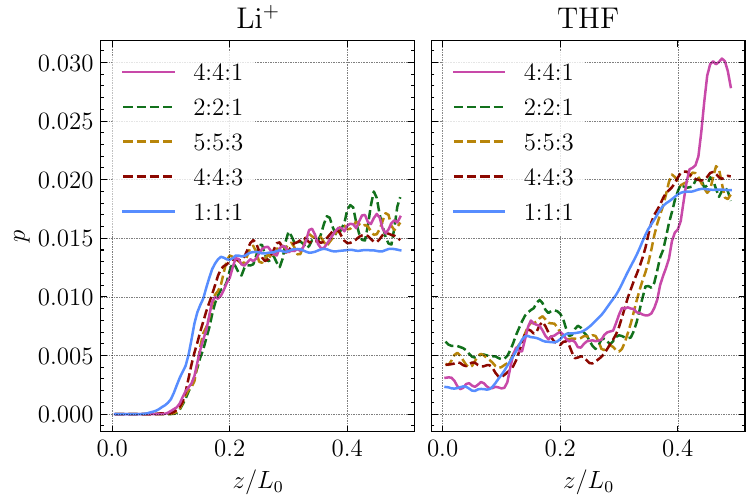}
 \caption{\label{fig:si_conc_profile_li_thf}Distribution of cations and THF along the normalized bilayer height $z/L_0$.
 The probability $p$ to find any given residue in a volume slice at $z$ is used instead of the mass density in order to account for volume changes due to the different amounts of THF.
 The data presented here directly reflects the mass density which is discussed in the main text (e.g.~the PEO domain is located at around $z/L_0 \in [0.2, 0.4]$).
 Notably, the \sysOneThfL{} system with a small amount of THF exhibits a higher relative abundance of THF in the central salt--rich layer.
 This indicates that, at low THF concentrations, the majority of THF is needed to compensate for the missing PEO coordination in the middle of the bilayer,
 whereas at higher concentrations, the majority of THF can distribute more freely.}
\end{figure}

\begin{figure}[H]
 \centering
 \includegraphics[width=0.8\textwidth]{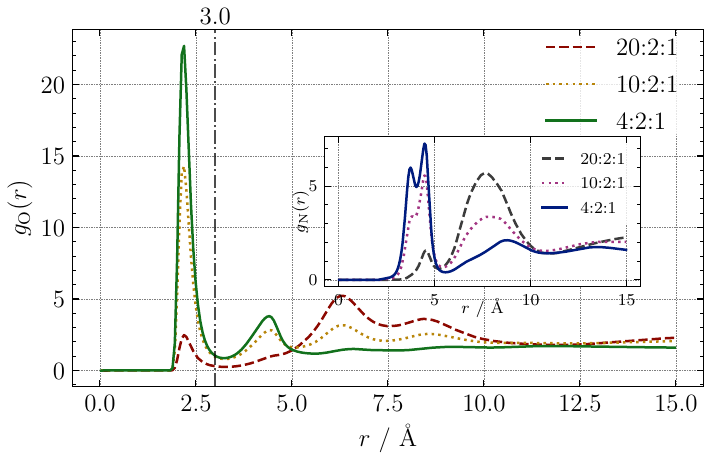}
 \caption{\label{fig:si_rdf_tfsi}RDF $g_{\ce{O}}(r)$ of TFSI oxygen and RDF $g_{\ce{N}}(r)$ of TFSI nitrogen (inset) around cations.}
\end{figure}

\begin{figure}[H]
 \centering
 \includegraphics[width=0.8\textwidth]{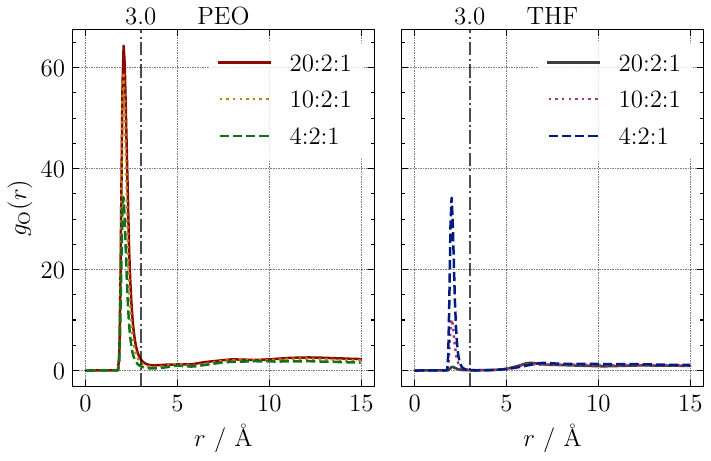}
 \caption{\label{fig:si_rdf_ether}RDF $g_{\ce{O}}(r)$ of PEO and THF oxygen around cations.}
\end{figure}

\begin{figure}[H]
 \centering
 \includegraphics[width=0.8\textwidth]{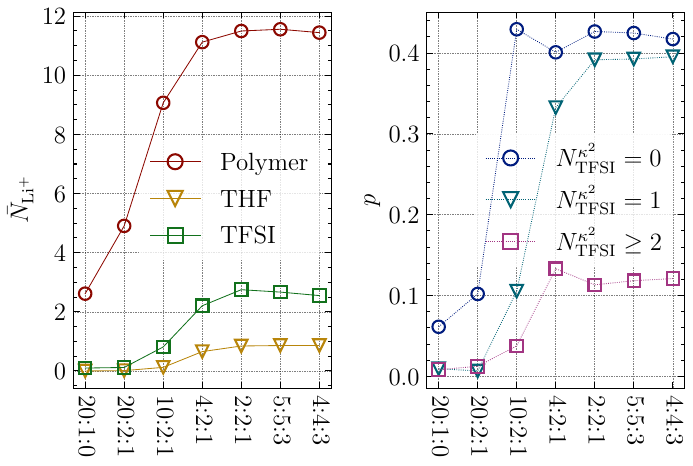}
 \caption{\label{fig:si_coord_res_li}Average number $\bar{N}_{\ce{Li+}}$ of cations that are coordinated to a molecule of a certain type (left).
 Fraction $p$ of cations that is either coordinated by monodentate \ce{TFSI} only ($N^{\kappa^2}_{\ce{TFSI}} = 0$) or coordinated by a certain number of bidentate anions (right).}
\end{figure}

\begin{figure}[H]
 \centering
 \includegraphics[width=0.8\textwidth]{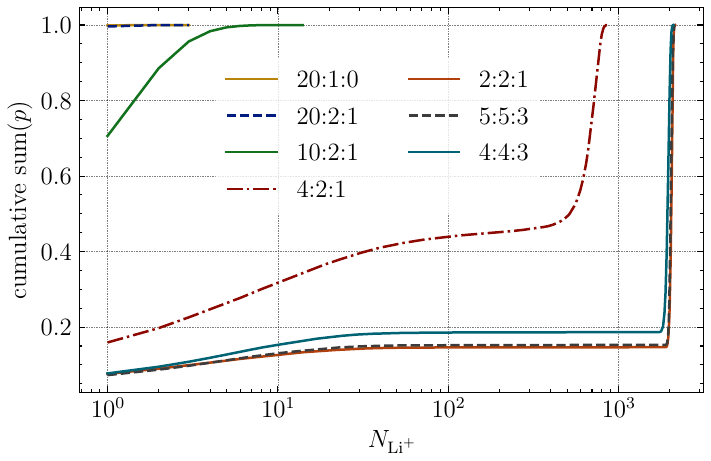}
 \caption{\label{fig:si_clusters}Cumulative probability to find a cation in an anion--cation cluster with a number of cations smaller or equal to $N_{\ce{Li+}}$.
 A cluster is defined as multiple cations bridged by anions.
 Cations bridged by PEO chains are not considered to be in the same cluster.}
\end{figure}

\begin{figure}[H]
 \centering
 \includegraphics[width=0.8\textwidth]{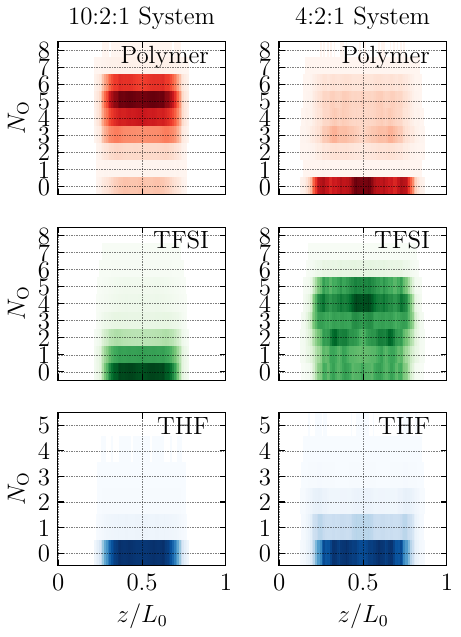}
 \caption{\label{fig:si_coord_profile}Comparison of the distribution of the number of coordinating oxygens in the \sysFive{} and \sysTwo{} systems along the normalized bilayer height $z/L_0$.}
\end{figure}

\begin{figure}[H]
 \centering
 \includegraphics[width=0.8\textwidth]{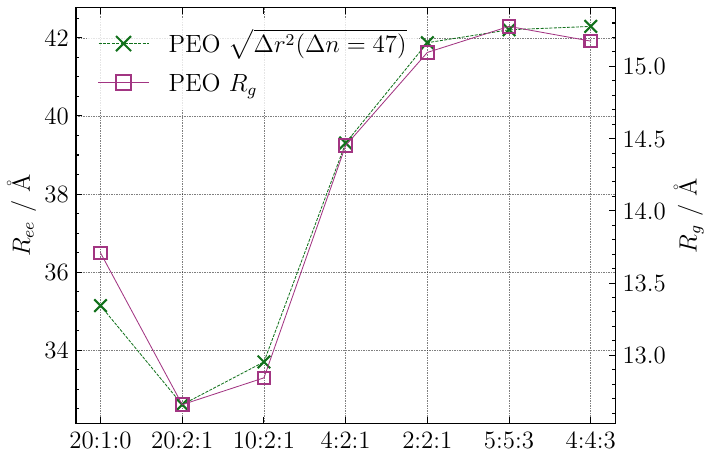}
 \caption{\label{fig:si_ree}Comparison between the end--to--end distance $R_{ee}$ and the radius of gyration $R_g$ of the PEO chains.
 The end--to--end distance $R_{ee}$ can be calculated from the squared distance between $\Delta n = 47$ interjacent bonds because the chains consist of 48 monomers.}
\end{figure}

\begin{figure}[H]
 \centering
 \includegraphics[width=0.8\textwidth]{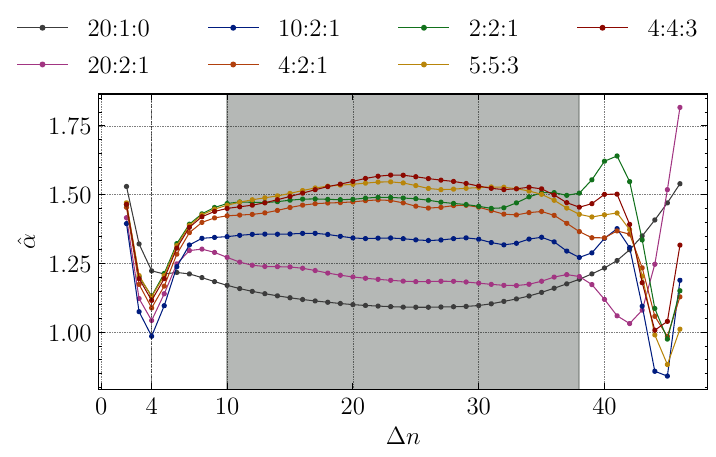}
 \caption{\label{fig:si_alpha_local}Local approximation $\hat{\alpha}$ of the stretching parameter.
 All systems exhibit a local minimum for $\Delta n = 4$ and a plateau in the interval $\Delta n \in [10, 38]$.}
\end{figure}

\begin{figure}[H]
 \centering
 \includegraphics[width=0.8\textwidth]{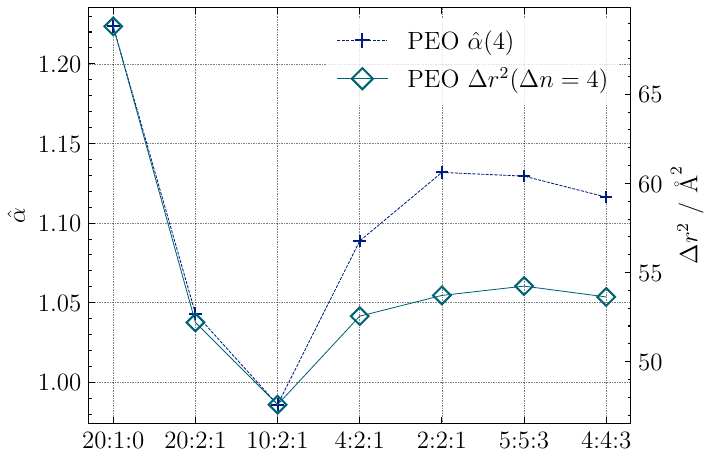}
 \caption{\label{fig:si_rsq_4}Comparison between the squared distance $\Delta r^2(4)$ and $\hat{\alpha}(4)$ of the PEO chains.}
\end{figure}

\begin{figure}[H]
 \centering
 \includegraphics[width=0.8\textwidth]{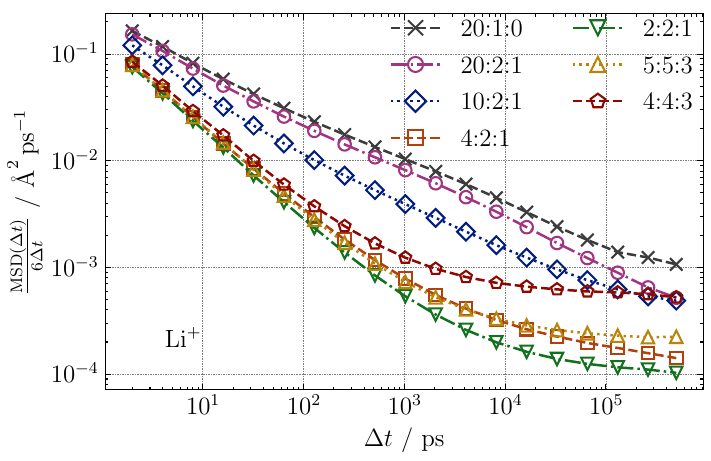}
 \caption{\label{fig:si_msd_li}Estimation of the cationic diffusion coefficient from the MSD according to the Einstein relation \cref{eq:einstein} in the main text.
 A plateau for long $\Delta t$ indicates truly diffusive motion.}
\end{figure}

\begin{figure}[H]
 \centering
 \includegraphics[width=0.8\textwidth]{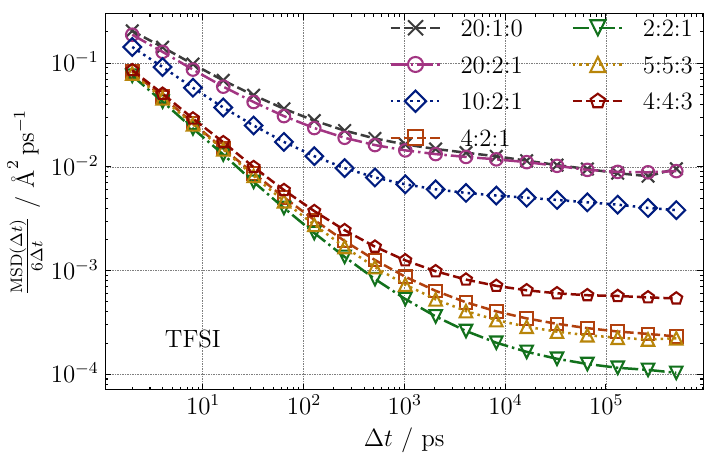}
 \caption{\label{fig:si_msd_tfsi}Estimation of the anionic diffusion coefficient from the MSD according to the Einstein relation \cref{eq:einstein} in the main text.
 A plateau for long $\Delta t$ indicates truly diffusive motion.}
\end{figure}

\begin{table}[H]
 \caption{\label{tab:si_diff_ions}Diffusion coefficients of the ions.}
 \begin{tabularx}{\columnwidth}{@{}>{\centering\arraybackslash}X>{\centering\arraybackslash}X>{\centering\arraybackslash}X@{}}
  \toprule
   System & $D_{\ce{Li+}}$ / \si{\square\angstrom\per\pico\second} & $D_{\ce{TFSI}}$ / \si{\square\angstrom\per\pico\second} \\ \hline
   \sysTwenty  & $9.83 \cdot 10^{-4}$ & $8.67 \cdot 10^{-3}$ \\
   \sysTen     & $4.31 \cdot 10^{-4}$ & $8.87 \cdot 10^{-3}$ \\
   \sysFive    & $4.53 \cdot 10^{-4}$ & $3.78 \cdot 10^{-3}$ \\
   \sysTwo     & $1.38 \cdot 10^{-4}$ & $2.28 \cdot 10^{-4}$ \\ \hline
   \sysOne     & $1.02 \cdot 10^{-4}$ & $1.02 \cdot 10^{-4}$ \\
   \sysOneThfA & $2.19 \cdot 10^{-4}$ & $2.15 \cdot 10^{-4}$ \\
   \sysOneThfB & $5.30 \cdot 10^{-4}$ & $5.38 \cdot 10^{-4}$ \\
  \bottomrule
 \end{tabularx}
\end{table}

\begin{table*}
 \caption{\label{tab:si_systems_salt_only}Systems with LiTFSI and THF but without polymer which are used to investigate the cation diffusion.
 Being equivalent to a polymer system means that the composition of the salt--rich layer has been determined in the polymer system,
 and the system described in this table has the same composition but lacks the polymer.}
 \begin{tabularx}{\textwidth}{@{}>{\centering\arraybackslash}X>{\centering\arraybackslash}X>{\centering\arraybackslash}X>{\centering\arraybackslash}X@{}}
  \toprule
   Equivalents & $N_{\text{LiTFSI}}$ & $N_{\text{THF}}$ & ${N_{\text{LiTFSI}}} / {N_{\text{THF}}}$ \\ \hline
   --- & 2400 & 3600 & 0.67 \\
   --- & 2400 & 2880 & 0.83 \\ \hline
   \sysOneThfB & 2400 & 2574 & 0.93 \\
   \sysOneThfA & 2400 & 1990 & 1.21 \\
   \sysOne & 2400 & 1540 & 1.56 \\ \hline
   --- & 2400 & 600 & 4.00 \\
  \bottomrule
 \end{tabularx}
\end{table*}

\begin{figure}[H]
 \centering
 \includegraphics[height=0.4\textheight]{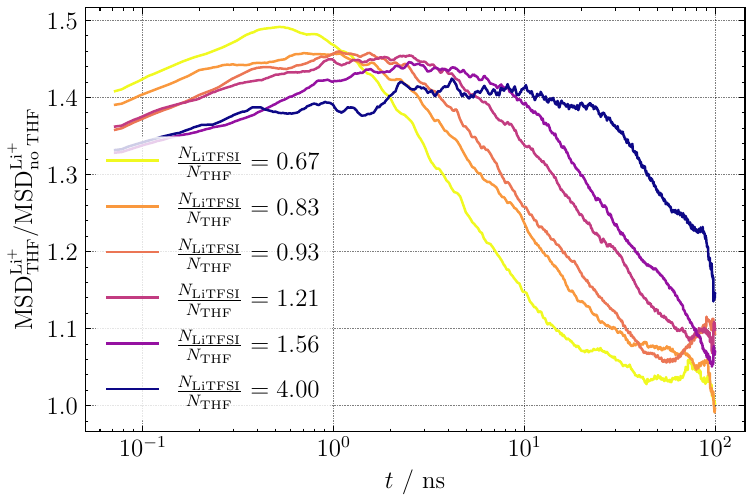}
 \caption{\label{fig:si_msd_li_thf_salt}Cationic mean squared displacement (MSD) determined in systems with LiTFSI and THF only (see \cref{tab:si_systems_salt_only}).
 The displacement of cations which are coordinated by THF at $t = 0$ is given relative to the displacement of cations with no THF in their first coordination shell.
 This is done to account for the system--wide plasticizing effect of THF (i.e.~an overall lower viscosity).}
\end{figure}

\begin{figure}[H]
 \centering
 \includegraphics[height=0.4\textheight]{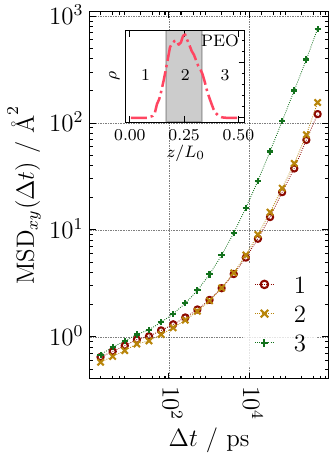}
 \caption{\label{fig:si_msd_layer}Lateral $\text{MSD}_{xy}$ for cations in the \sysOneThfB{} system.
 Illustration of the three layers in relation to the PEO domain (shaded region in inset).}
\end{figure}

\begin{table}[H]
 \caption{\label{tab:si_diff_ions_layer}Lateral diffusion coefficients $D_i$ of the ions in the distinct layers in the \sysOneGroup{} systems.
 The standard deviation $\sigma^D_i$ was computed by randomly dividing the ions into three groups of equal size and comparing the group's respective diffusion coefficients.}
 \begin{tabularx}{\columnwidth}{@{}>{\centering\arraybackslash}X>{\centering\arraybackslash}X>{\centering\arraybackslash}X>{\centering\arraybackslash}X>{\centering\arraybackslash}X>{\centering\arraybackslash}X@{}}
  \toprule
   System & Layer & $D_{\ce{Li+}}$ / \si{\square\angstrom\per\pico\second} & $D_{\ce{TFSI}}$ / \si{\square\angstrom\per\pico\second}
        & $\sigma^{D}_{\ce{Li+}}$ / \si{\square\angstrom\per\pico\second} & $\sigma^{D}_{\ce{TFSI}}$ / \si{\square\angstrom\per\pico\second} \\ \hline
\multirow{2}{*}{\sysOne}     & 2 & $5.36 \cdot 10^{-5}$ & $7.48 \cdot 10^{-5}$ & $4.34 \cdot 10^{-6}$ & $4.88 \cdot 10^{-6}$ \\
                             & 3 & $1.50 \cdot 10^{-4}$ & $1.31 \cdot 10^{-4}$ & $3.23 \cdot 10^{-6}$ & $1.91 \cdot 10^{-6}$ \\ \hline
\multirow{2}{*}{\sysOneThfA} & 2 & $7.12 \cdot 10^{-5}$ & $1.03 \cdot 10^{-4}$ & $5.75 \cdot 10^{-6}$ & $1.59 \cdot 10^{-6}$ \\
                             & 3 & $3.74 \cdot 10^{-4}$ & $3.38 \cdot 10^{-4}$ & $3.49 \cdot 10^{-5}$ & $4.36 \cdot 10^{-6}$ \\ \hline
\multirow{2}{*}{\sysOneThfB} & 2 & $1.74 \cdot 10^{-4}$ & $2.27 \cdot 10^{-4}$ & $2.35 \cdot 10^{-6}$ & $7.02 \cdot 10^{-6}$ \\
                             & 3 & $1.00 \cdot 10^{-3}$ & $9.65 \cdot 10^{-4}$ & $2.85 \cdot 10^{-5}$ & $4.46 \cdot 10^{-5}$ \\
  \bottomrule
 \end{tabularx}
\end{table}